\documentclass[prb,preprint,showpacs,amssymb,amsfonts]{revtex4}
\usepackage{graphicx}

\newcommand{\del}{\partial}

\begin{document}

\title{Universality Classes of Metal-Insulator Transitions in Strongly Correlated Electron Systems and Mechanism of High-Temperature Superconductivity}
\author{Masatoshi Imada}
\affiliation
{
Institute for Solid State Physics, 
University of Tokyo, 5-1-5, Kashiwanoha, Kashiwa, Chiba 277-8581
%and \\
%PRESTO, Japan Science and Technology Agency
}

\begin{abstract}
We study three regimes of the Mott transitions characterized by classical, marginally quantum and quantum. In the classical regime, the quantum degeneracy temperature is lower than the critical temperature of the Mott transition, $T_c$, below which the first-order transition occurs. The quantum regime describes the $T_c=0$ boundary of the continuous transition. The marginal quantum region appears sandwiched by these two regimes.  The classical transition is described by the Ising universality class. However, the Ginzburg-Landau-Wilson scheme breaks down when the quantum effects dominate. The marginal quantum critical region is categorized to a new universality class, where the order parameter exponent $\beta$, the susceptibility exponent $\gamma$ and the field exponent $\delta$ are given by $\beta=d/2, \gamma=2-d/2$ and $\delta=4/d$, respectively, with $d$ being the spatial dimensionality. It is shown that the transition is always at the upper critical dimension irrespective of the spatial dimensions. Therefore, the mean-field exponents and the hyperscaling description become both valid at any dimension.  The obtained universality classes agree with the recent experimental results on the Mott criticality in organic conductors such as $\kappa$-(ET)$_2$Cu[N(CN)$_2$]Cl and transition metal compounds such as V$_2$O$_3$. The marginal quantum criticality is characterized by the critically enhanced electron-density fluctuations at small wavenumber. The characteristic energy scale of the density fluctuation extends to the order of the Mott gap in contrast to the spin and orbital fluctuation scales and causes various unusual properties.   The mode coupling theory shows that the marginal quantum criticality further generates non-Fermi-liquid properties in the metallic side.  The effects of the long-range Coulomb force in the filling-control Mott transition are also discussed.  A mechanism of high temperature superconductivity emerges from the density fluctuations at small wavenumber inherent in the marginal quantum criticality of the Mott transition. The mode coupling theory combined with the Eliashberg equation predicts the superconductivity of the $d_{x^2-y^2}$ symmetry with the transition temperature of the correct order of magnitude for the realistic parameters for the cuprate superconductors.   Experimental results on the electron differentiations established in the angle-resolved photoemission experiments are favorably compared with the present prediction.  The tendency for the spatial inhomogeneity is a natural consequence of this criticality.   
%\pacs{73.20.Qt, 71.10.Ca, 71.30.+h, 71.10.Fd, 73.21.-b, 05.10.-a} 
%\kword{Ginzburg-Landau-Wilson theory, quantum phase transition, universality class, Mott transition, two-dimensional Hubbard model, high-Tc superconductivity, phase separation, non-Fermi liquid, electron differentiation, organic conductors,transition metal oxides
\end{abstract}
%\sloppy
\maketitle

\date{January 6, 2005}
%%%%%%%%%%%%%%%%%%%%%%%%%%%%%%%%%%%%%%%%%%%%
%% MAINMATTER
%%%%%%%%%%%%%%%%%%%%%%%%%%%%%%%%%%%%%%%%%%%%

\section{Introduction}

Mott transition belongs to one of the metal-insulator transitions ubiquitous in various compounds~\cite{RMP}.  Physical properties of Mott transition and its nature are a long-standing subject of research with many controversial issues.  The problem has first been postulated in 1937 by Peierls and Mott~\cite{PeierlsMott}. Although the band theory of metals and insulators established soon after the foundation of quantum mechanics is quite successful, it was pointed out by de Boer and Verway~\cite{deBoer} that insulating behaviors of many transition metal compounds as NiO cannot be explained by a simple band picture, because the bands are partially filled.  Peierls pointed out a crucial role of electron correlations as the mechanism of the insulating behavior. Mott developed this idea and introduced a concept which we nowadays call the Mott insulator~\cite{Mott,Mott2}.  Since then, it has been recognized for a long time that the Mott insulator stabilized by the electron-electron Coulomb interaction and metallic states near it provide us with fruitful physics with various novel concepts.  This has become very popular after the discovery of the high-$T_c$ cuprate superconductors, which was indeed discovered in doped Mott insulators~\cite{Bednorz}.
However, relationship of this fruitful outcome to the nature of the Mott transition itself has not been fully clarified.

The Mott transition can be realized basically by two routes: Bandwidth-control and filling-control.
In the first route, the bandwidth is controlled relative to the amplitude of the local electron-electron interaction by keeping the electron density fixed at a commensurate value (namely, the electron density per unit cell is kept at an integer value). 
This route may be experimentally realized by applying pressure or by substituting elements with a different ionic radius and the same valence.  In this route, the overlap of the wavefunctions between neighboring electronic atomic orbitals forming conduction electron bands is controlled. 
In the second route, electron filling is changed from the Mott insulator.  This is typically realized by substituting with elements, which have a different valence from the substituted elements in the reservoir structure of the Mott insulator.

\begin{figure}
\begin{center}
\includegraphics[width=7cm,clip]{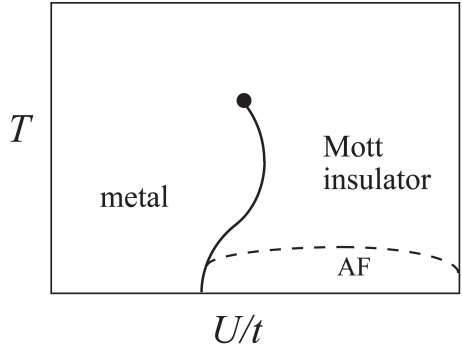}
\end{center}
\caption{Schematic and typical phase diagram of metal and Mott insulator in the parameter space of temperature $T$ and the interaction strength scaled by the transfer, $U/t$. The solid curve shows the first-order Mott transition with the critical end point illustrated by the solid circle. In many cases, the antiferromagnetic (AF) phase appears at low temperatures as is illustrated by the broken phase boundary line.}
\label{Fig0}
\end{figure}

Originally, Mott has considered only the first route, the bandwidth-control transition. In this category, there exist several typical Mott transitions including those observed in V$_2$O$_3$, $R$NiO$_3$ and $\kappa$-ET type organic compounds, where $R$ represents a rare earth element.  Many of them show first-order transitions between the Mott insulator and metals at low temperatures~\cite{RMP}.  In many cases, some magnetic and/or orbital order exist at low temperature of the Mott insulator phase.  Although sometimes concurrent magnetic transitions occur with the Mott transition at low temperatures, in the typical phase diagram illustrated in Fig.~\ref{Fig0}, the first-order Mott transitions occur even at high temperatures, where the magnetic and orbital orders are not involved.  This typical phase diagram clearly indicates that the Mott transition is inherently independent of the magnetic and orbital-ordering transitions. The statement that the mechanism of the Mott transition is primarily independent of the symmetry breaking of spins or orbitals is also corroborated by the existence of a quantum spin liquid phase in the Mott insulator recently found numerically~\cite{Kashima1,Morita} as well as experimentally~\cite{KanodaSpinliquid,Fukuyama,Ishimoto}. The Mott transition between a quantum spin-liquid and a metal does not accompany magnetic transitions. In fact, as we see in this paper, the Mott transition is not driven by the spin or orbital degrees of freedom but by the density degrees of freedom.  

Mott argued that the long-range part of the Coulomb interaction is necessary to reproduce the first-order transition~\cite{Mott1storder}.  The argument by Mott was the following: We can identify the Mott insulating phase as that where the two electrons sitting on the same atomic orbitals (we call it a doublon) and an empty site (we call it a holon) make a bound state.  Metals are characterized by the phase where the bound state disappears.  In terms of the binding of the doublon and holon, the binding energy is controlled by the screening of the attractive Coulomb interaction between the doublon and holon.  Since the screening relies on other doublons and holons moving in between, the screening becomes rapidly poor with decreasing density of free doublons and holons.  Then at some threshold concentration of free doublons and holons, the screening becomes too weak to keep a doublon and a holon free and as a consequence of this feedback, they suddenly form a bound state, which leads to a first-order transition to a Mott insulating state.  Another argument for the origin of the first-order transition emphasizes couplings to lattice distortion.  When the metal is stabilized, the lattice constant diminishes further to gain the kinetic energy of electrons, which in general strongly favors the first-order transition through the electron-lattice coupling.  

\begin{figure}
\begin{center}
\includegraphics[width=8cm,height=7cm,clip]{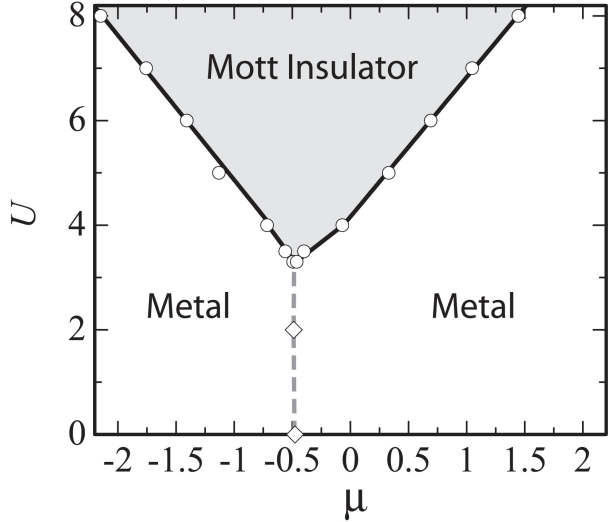}
\end{center}
\caption{ Phase diagram of Mott insulator (shaded area) and metal for two-dimensional Hubbard model on a square lattice with the nearest-neighbor transfer $t$ and the second-neighbor transfer $t'$ in the plane of the local interaction $U$ and the chemical potential $\mu$.  The energy unit is $t=1$ and $t'$ is taken at 0.2.  The bandwidth-control route is realized across the bottom corner of the Mott phase boundary along the gray dashed line, whereas the filling-control transition occurs across the solid lines.}
\label{phasediagram.eps}
\end{figure}

\begin{figure}
\begin{center}
%$$ \psboxscaled{600}{SVOLDADispnew.eps} $$
\includegraphics[width=8cm,height=6cm]{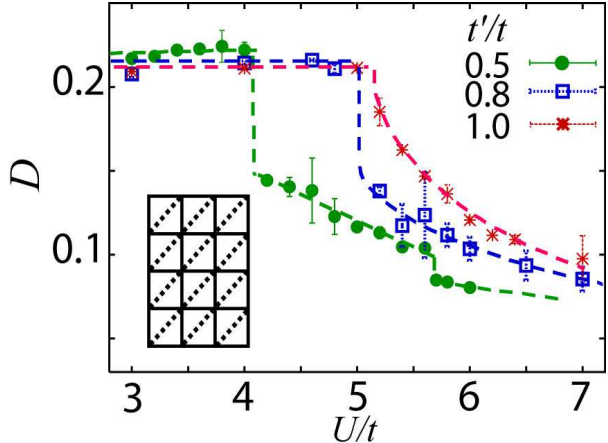}
\end{center}
\caption{Double occupancy $D$ at $T=0$ as a function of $U/t$ for three choices of parameters, $t'/t=0.5, 0.8$,and 1.0 for the anisotropic triangular lattice structure depicted in the inset with the nearest neighbor transfer $t$ (solid bond) and the next-nearest-neighbor transfer $t'$ (broken bond). Jumps of $D$ at $U/t=4.1,5.0$ and 5.2 for $t'/t=0.5,0.8$ and 1.0, respectively show the first-order Mott transitions. Curves are guide for the eyes}
\label{Djump.eps}
\end{figure}

However, recent detailed numerical studies on the Hubbard model on the square lattice overturned these speculations~\cite{Kashima1,Morita,WatanabeGPIRG}: Although the long-range Coulomb force and the coupling to the lattice may have some roles in the realistic Mott transition, even the Hubbard model with only the local onsite interaction without any coupling to lattice distortions is enough to reproduce the first-order Mott transition.  The $N$-site Hubbard model is defined as 
\begin{eqnarray}
{\cal H} & = &{\cal H}_t + \sum_i H_{Ui} -\mu M N \label{1.1} \\
{\cal H}_t & = & -\sum_{\langle
ij\rangle}t_{ij}(c^{\dagger}_{i\sigma}c_{j\sigma}+h.c.) \label{1.2} 
\end{eqnarray} 
and
\begin{equation}
{\cal H}_{Ui} = U (n_{i\uparrow}-\frac{1}{2})(n_{i\downarrow}-\frac{1}{2})\label{1.3}.
\end{equation}  
%represents this competition in a simplest way.  
Here, 
$M  \equiv  \sum_{i\sigma} n_{i\sigma}/N$ and $n_{i\sigma}=c^{\dagger}_{i\sigma}c_{i\sigma}$ with the creation (annihilation) operator $c^{\dagger}_{i\sigma} (c_{i\sigma})$ of an electron at the site $i$ with the spin $\sigma$.  
The chemical potential is $\mu$ and $U$ is the onsite Coulomb repulsion. 
The phase diagram of the Hubbard model at zero temperature $T=0$ on the square lattice with the nearest-neighbor and next-nearest-neighbor transfers $t$ and $t'$, respectively, studied by the path-integral renormalization group method~\cite{ImadaKashima,KashimaImadaPIRG} is shown in the plane of $U/t$ and the chemical potential $\mu$ at zero temperature in Fig.~\ref{phasediagram.eps}~\cite{WatanabeGPIRG}. The filling-control transition occurs across the edge of the boundary in Fig.~\ref{phasediagram.eps}, while the bandwidth-control transition is realized through the corner of the phase boundary at the bottom.  The first-order transition through the bandwidth-control route is indicated by the jump of the averaged doublon density $D\equiv \langle n_{i\uparrow}n_{i\downarrow}\rangle=\sum_in_{i\uparrow}n_{i\downarrow}/N $ as in Fig.~\ref{Djump.eps} for the case of the anisotropic triangular lattice~\cite{Morita}. When the first-order transition takes place at zero temperature, its jump decreases with raising temperatures and closes at the critical endpoint.  In fact, the Hubbard model on this anisotropic triangular lattice is a relevant effective model for the $\kappa$-ET type organic compound and the first-order Mott transition with the finite-temperature critical end point was observed experimentally~\cite{Fournier,Kanoda,Lefebvre}. 

Another remarkable feature in Fig.~\ref{Djump.eps} is that the jump in $D$ decreases with increasing the next nearest neighbor transfer $t'$. This means that the so-called frustration effects reduce the first-order jump and drives to more continuous type transitions.  In fact, recent experimental studies on pyrochlore compounds~\cite{Tokura} appear to show a continuous Mott transition by the bandwidth control with a signature of the Anderson localization in the vicinity of the boundary.  We note that the pyrochlore lattice has the fully frustrated structure, where the antiferromagnetic order is severely suppressed.  In addition, recent studies on $\kappa$-(ET)$_2$Cu$_2$(CN)$_3$ suggest that the critical temperature of the Mott transition becomes lower for the more frustrated structure, namely for compounds with larger effective $t'$ in the corresponding theoretical model.~\cite{Kanoda3}   All of these consistently suggest that the order of the Mott transition may be systematically controlled from the first order to continuous.

When the filling is controlled, the first-order transition appears as the phase separation.  Experimentally, the existence of the phase separation or the electronic inhomogeneity is a controversial issue as we discuss later.  From theoretical side, numerical studies on the Hubbard model show the marginal result, where the phase separation does not occur while the charge susceptibility shows a critical divergence at the Mott transition.~\cite{FurukawaImada0,FurukawaImada,FurukawaImada2} In the terminology in this paper, we use the charge susceptibility and the density susceptibility as the same quantity $d\langle M \rangle /d\mu$ for the electrons with charge.    

Since the theoretical and experimental results suggest the controllability of the order of the Mott transition and its critical temperature, it is desired to understand the whole feature of the Mott transition from the universality classes of the finite-temperature critical point to the zero-temperature critical phenomena on the same grounds.  We will show in this paper that three regimes of the Mott transition exist. One is the classical transition at a high temperature, which is described by the Ising universality class of the critical point accompanied by the first-order transition below the critical temperature. The second is the quantum transition, where the transition appears only at zero temperature, and the density susceptibility remains finite for the spatial dimension $d \ge 2$.  The third regime is the marginally quantum one, which emerges at the crossing point of the classical and quantum transitions. The marginally quantum regime is characterized by the diverging density susceptibility at small wavenumber for $d \ge 2$ at low temperatures. In the second and third regimes, the conventional scheme of the Ginzburg-Landau-Wilson theory does not apply. 

The new universality class at the marginally quantum transition has a deep consequence on the induced non-Fermi-liquid behavior in the metallic side, electron differentiation in the momentum space and the sensitivity toward electronic inhomogeneity.
One of the most remarkable consequences is the superconductivity emerging from this marginal quantum Mott criticality. We show that the high temperature superconductivity of the $d_{x^2-y^2}$ symmetry is obtained under the realistic choice of the parameter values for the cuprate superconductors, where the density (charge) fluctuations at small wavenumber play the crucial role for the Cooper pairing.  The energy scale of the fluctuation is characterized by the Mott gap, which can be by far larger than the energy scale of magnetic and orbital fluctuations for the filling-control transition. This solves many puzzling experimental results in strongly correlated electron systems particularly in transition metal compounds.  

A part of the discussions in this article is already given~\cite{Imada2004,Imada2005}. We summarize the previous results and further extend the discussion on the quantum Mott criticality and its consequences in greater detail. 
In particular, detailed analyses on the breakdown of the Ginzburg-Landau-Wilson scheme for the quantum Mott transition are presented together with the scaling analysis. The validity of the mean-field exponents and the compatibility with the hyperscaling description are discussed in detail.   Two-dimensional systems are especially analyzed and are compared with the experimental results for the organic compounds and the cuprate superconductors.  Basic finite temperature effects are also obtained and discussed in connection with the experimental results.  An important issue for the filling-control transition is the effects of the long-range Coulomb interaction.  We discuss how the present results are modified in the presence of the long-range interaction and also discuss the experimental relevance.
The non-Fermi-liquid properties and the anisotropic Cooper pairing originating from the nonperturbative enhancement of the density susceptibility at a small wavenumber is an important subject we study in this article in detail. 

In Sec. II, we summarize the conventional Ginzburg-Landau-Wilson scheme of the Mott transition. In Sec. III, the nature of the quantum Mott transition is examined with emphasis on the breakdown of the Ginzburg-Landau-Wilson scheme.  Section IV is devoted to the resultant non-Fermi-liquid behavior expected in the metallic side of the critical region of the Mott transition.  In Sec. V we discuss the mechanism of high-temperature superconductivity arising from the marginal quantum Mott criticality. Section VI concludes and summarizes the paper.

\section{Conventional Ginzburg-Landau-Wilson Scheme}
Recently, the critical endpoint of the first-order transition line of the Mott transition in Fig.~\ref{Fig0} has been a subject of intensive studies. In case of V$_2$O$_3$, from the detailed study of the conductance, it has been suggested that the criticality of the transition follows the Ising-type universality class~\cite{Limelette}. In an organic conductor of $\kappa$-ET salt, the diverging electronic compressibility at the critical end point has been probed by the ultrasound velocity~\cite{Fournier}. 

In prior to these experimental studies, several theoretical studies have focused on the nature of the transition. 
The Mott transition by itself does not change any symmetry. Therefore, from theoretical point of view, it has an analogy with the text-book gas-liquid transition, which is known to be equivalent to the ferromagnetic transition in the Ising model under magnetic fields.  The first-order metal-insulator transition corresponds to the Ising transition between spin-up and down phases taking place with switching the direction of magnetic fields below the critical temperature.  In fact, Castellani~\cite{Castellani} has discussed the Ising nature of the first-order Mott transition by extending the Blume-Emery-Griffiths model~\cite{BEG} for the phase separation of $^3$He-$^4$He mixture.  It has further been considered in the dynamical mean-field theory by Kotliar et al.~\cite{Kotliar,Kotliar2}, where the low-energy part of the single-particle Green's function appears to follow Ginzburg-Landau scheme in accordance with the mean-field theory of the Ising model.   

In the Ginzburg-Landau-Wilson scheme for the Ising-type transition, the free energy may be expanded by the spatially dependent scalar order parameter $X(r)$ integrated over space coordinate $r$ as
\begin{equation}
F = \int dr [\frac{1}{2}a_0(T-T_c)X(r)^2+\frac{1}{4}bX(r)^4-\mu X(r)]
\label{1.4} 
\end{equation} 
near the critical temperature $T_c$ with $a_0$ and $b$ being positive constants.  In the Ising model, $X$ is indeed the order parameter, namely, the magnetization $m$.  In the mapping to the gas-liquid transition, $X$ is interpreted as the density of particles, $n$, measured from the critical density.  At the critical temperature $T_c$ of the gas-liquid transition, the uniform density susceptibility $\chi_n=[d^2F/dn^2]^{-1}$ diverges.  

When it is further extended to the mapping to the Mott transition, for the filling-control transition, $X$ is identified indeed as the electron doping concentration $X$ measured from the critical density at the critical point of the Mott transition~\cite{Imada2004}.
This is a natural consequence, because, in the filling-control transition, the control parameter is the chemical potential, which is conjugate to the carrier density.   

When the bandwidth is controlled, the control parameter is $U/t$ in the Hubbard model and the conjugate quantity to $U$ is the doulon density $D$.  Therefore, the order parameter $X$ in this case is the doublon density $D$, which indeed jumps at the first-order transition as in Fig.~\ref{Djump.eps}. We note that in this case the holon density should be the same as the doublon density, because the holon and doublon densities must be the same to keep the density for the route of the bandwidth-control transition. 

To describe the both types of the transitions, we take the natural order parameter as $\zeta$, where $\zeta=X$ for the filling-control transition and $\zeta=D$ for the bandwidth-control transition.
Namely, we take 
\begin{equation}
F = \int dr [\frac{1}{2}a_0(T-T_c)\zeta(r)^2+\frac{1}{4}b\zeta(r)^4-\mu_{\zeta} \zeta(r)].
\label{1.4-2} 
\end{equation} 
Here, $\mu_{\zeta}$ is $-U$ for the bandwidth-control transition whereas is the chemical potential $\mu$ conjugate to the doping concentration $X$ for the filling-control transition.
The Ising universality is resulted from this Ginzburg-Landau-Wilson functional~\cite{Goldenfeld}. 
When the Ising universality is correct, 
we obtain the critical susceptibility exponent defined by $\chi_c\equiv [\partial^2 F /\partial \zeta^2]^{-1} \propto (T-T_{c})^{-\gamma}$ with $\gamma_=7/4$ for two-dimensional systems, $d=2$ and $\gamma\sim 1.24$ for $d=3$. The order parameter exponent defined below $T_c$ as $\langle \zeta \rangle \propto |T-T_c|^{\beta}$ obtained from $\partial F/\partial \zeta=0$ at $\mu_{\zeta}=0$ satisfies $\beta=1/8$ and $\beta\sim 0.325$ for $d=2$ and 3, respectively.  The exponent with varying $\mu_{\zeta}$ defined as $\langle \zeta \rangle\propto \mu_{\zeta}^{1/\delta}$ at $T=T_c$ is given by $\delta=15$ and 4.8 for $d=2$ and 3, respectively.  These exponents for $d=3$ were indeed claimed to be observed at the critical point of the bandwidth-control Mott transition for V$_2$O$_3$~\cite{Limelette}. 

In the Ising-transition picture~\cite{Goldenfeld}, the transition is characterized by these simple exponents with the hyperscaling assumption being satisfied below the upper critical dimension $d_u=4$.
The system has a single length scale $\xi$ which diverges at $T=T_c$.  In the present context, $\xi$ expresses the density correlation length or doublon density correlation length.  

\section{Quantum Mott Criticality}
\subsection{General remark on quantum effect}
A nontrivial question arises when the critical temperature of the Mott transition, $T_c$ can be lowered.  
With lowering of $T_c$, how do the quantum effects emerge?  
If $T_c$ becomes zero, then one might naively expect that conventional quantum critical phenomena would appear. 
%The universality of such a quantum phase transition cannot be the same Ising type, because the Ising transition is genuinely classical one and cannot describe the quantum criticality.  
A naive expectation would be that the transition might be described by the Ising universality class in $d+1$ dimensions, where the additional one dimension emerges from the dimension in the imaginary time in the path integral formalism.  It turns out later that this is not the case. In any case, the criteria for the existence of the non-negligible quantum effect should be determined from the existence of the Fermi degeneracy of the electrons.  When the Fermi degeneracy temperature becomes comparable or higher than $T_c$, the Mott transition has to be treated fully quantum mechanically.

One may argue that even when the bare Fermi temperature is high, the effective Fermi temperature would be suppressed near the Mott critical point, because, at the continuous transition point to the insulator, the Fermi degeneracy temperature should be zero. However, the Fermi degeneracy may still coexist with the critical fluctuation in the metallic side near $T_c$. In fact, even when $T_c$ is zero, the Fermi degeneracy always appears at temperatures sufficiently close to zero, if the parameter infinitesimally deviates from the critical point. Then the quantum effect should become relevant when one approaches the transition point by keeping the temperature sufficiently low.  
  
% 経路積分による考察、動的臨界指数の重要性、
In the quantum region, we have to consider quantum dynamics.  This can be done by considering the path-integral formalism, where the imaginary time direction must be additionally considered in addition to the real spatial dimension. The time scale $\omega^{-1}$ diverges as $\omega^{-1} \propto \xi^{z}$ in addition to the divergence of the spatial correlation length $\xi$.   The quantum dynamics is characterized by the dynamical exponent $z$. We note here that the Mott transition can be characterized by two different dynamical exponents in principle. This is because at the Mott transition, the single-particle spectra given by the single-particle Green's function and the two-particle correlations represented by the density (charge) correlation functions both have singular behaviors with diverging time scale, while these two may in principle follow different scalings.  Therefore we can define two dynamical exponents $z$ and $z_t$ for single-particle and two-particle spectra, respectively.  We will show below that these two coincide each other.

% 臨界温度がゼロのプロトタイプの考察、純理論的に面白いだけでなく、ＦＣＭＴ数値計算も支持する
% 引力が弱くて量子領域に入ってきたとき、,,根本的問題
Here, we discuss how the quantum effect alters the transition by assuming the region where $T_c$ is low or even zero.   
When $T_c$ becomes low, the Ginzburg-Landau expansion tells that the charge fluctuation becomes diverging accompanied by the quantum degeneracy, which becomes beyond the scope of the form (\ref{1.4-2}).  One might expect that the quantum region could be described by the Ising universality in $d+z$ dimensions. We will show that this does not apply and the quantum effect is more profound.

\subsection{What is different from the conventional quantum critical phenomena?}
In contrast to phase transitions with simple spontaneous symmetry breakings, the transitions from metals to the band insulators and the Mott insulators have no spontaneous symmetry breaking by themselves. Therefore as in the gas-liquid transition, the metals and insulators have no clear distinction at nonzero temperatures, if the first-order transition would be absent.   However, at zero temperature, insulators are always clearly distinguished from metals by the vanishing conductivity.  Among various types of insulators, band insulators and Mott insulators both have clear distinction from metals by vanishing Drude weight and vanishing charge susceptibility (compressibility) at zero temperature~\cite{RMP,Imada1995}.  The Drude weight is the stiffness to the twist of the phase of the wavefunction in the spatial direction, while the charge susceptibility is the stiffness to the twist in the temporal direction in the path-integral formalism~\cite{Kohn, Fisher, RMP}. Both of the two quantities have nonzero values only in metals.   Therefore, metals under the perfectly periodic potential of ions are regarded as a state where symmetry of the phase of the spatially-extended electron wavefunction is broken.  Then, at zero temperature, insulators cannot be adiabatically continued to a metal. Two phases have to be clearly separated by a phase boundary. 

\begin{figure}
\begin{center}
\includegraphics[width=8cm,height=5cm]{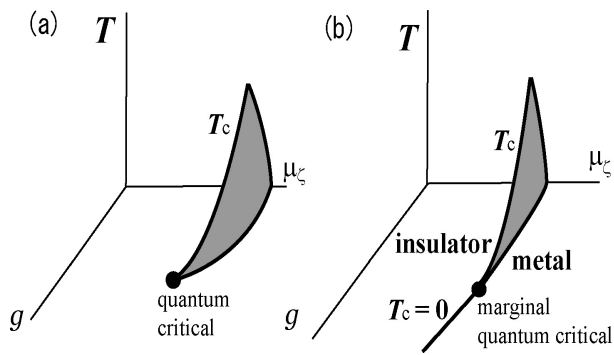}
\end{center}
\caption{Schematic phase diagram of conventional quantum critical phenomena (a) and the Mott transition (b) in the parameter space of the control parameter $\mu_{\zeta}$ conjugate to the order parameter, the parameter $g$ to control the quantum fluctuation and temperature $T$. In (b), the $T_c=0$ {\it boundary} continues as a line beyond the {\it marginal quantum critical point}. The shaded areas are the first-order phase boundaries. }
\label{Fig1}
\end{figure}

From these facts, we notice that the Mott transition at zero temperature may have a quite different universality class.  In fact, if one could lower $T_c$ in Eq.~(\ref{1.4-2}) by controlling some microscopic parameter, one may also expect that it could pass through zero and even to a negative temperature. This implies that the transition would become quantum critical and then the Mott transition would disappear as in the conventional scenario of the emergence of quantum critical phenomena as schematically illustrated in Fig.~\ref{Fig1}(a).  However, we have seen above that this cannot happen because of the clear distinction between metals and insulators at $T=0$.  This by itself indicates that the Ginzburg-Landau-Wilson scheme has to break down when $T_c$ becomes zero. 
When $T_c$ is lowered to zero, and if one tries to drive the control parameter further to lower $T_c$, it in reality keeps $T_c$ at zero, namely the quantum transition at $T=0$ continues as we see in Fig.~\ref{Fig1}(b). We call this continuation line, the $T_c=0$ {\it boundary}.  When $T_c$ becomes just zero from nonzero values, we call this point the {\it marginal quantum critical point}, which is indicated by the solid circle in Fig.~\ref{Fig1}(b). 

We have discussed already that the first-order transitions with the critical end point are indeed found in experiments.  We have also discussed and will discuss later that the metal-insulator transition through the $T_c=0$ boundary appears to exist.  Then the marginal quantum critical region is the crossing point of these two regions.  We will clarify that many strongly correlated systems including the high-$T_c$ cuprates may be located in this marginal quantum critical region.  Numerical results of the filling-control transition in the two-dimensional Hubbard model indeed suggest a continuous transition at zero temperature with the diverging charge susceptibility (density susceptibility)~\cite{FurukawaImada0,FurukawaImada,FurukawaImada2}, which is consistent with what is expected in the marginal quantum critical region.  Therefore, this quantum criticality is certainly a realistic possibility.
From experimental point of view, the quantum parameter $g$ may be controlled by the lattice structure, particularly, by the geometrical frustration as already discussed in \S I.  The control of the frustration parameter was actually achieved by the choice of anions in $\kappa$-type ET compounds while uniaxial pressure may also be used to control the frustration effects in general.

\subsection{Single-particle quantum dynamics}
\label{III.C}
The metallic phase except for one-dimensional systems has the adiabatic continuity with the Fermi liquid.  Therefore, low-energy part of the free energy can be described by the fermionic operators of renormalized single particles, where the higher-order terms are renormalized to the single-particle coefficient. 
In the insulating side as well, single-particle Green's function $G$ describes the charge dynamics and may be given from a quasiparticle description  
with a gap $\Delta_c(q)$ as
\begin{equation}
G(q,k,\omega_n)^{-1}=-i\omega_n + E(q,k)-\mu,
\label{Disp1}
\end{equation}
where the Matsubara frequency is $\omega_n$,
and $E(q,k)=\pm \sqrt{\Delta_c(q)^2 + \varepsilon(q,k)^2}$ with the bare dispersion $\varepsilon$.
In this expression, $k$ is the momentum coordinate perpendicular to the locus of $\varepsilon(q,k)=0$  and $q$ denotes that parallel to $\varepsilon(q,k)=0$. 
%with a gap $\Delta_c$ as
%\begin{equation}
%S=\sum_{i,n} \phi^* (-i\omega_n + \Delta_c + E(q,k))\phi,
%\label{Disp1}
%\end{equation}
%where the Matsubara frequency is $\omega_n$,
Here $E(q,k)$ is assumed to satisfy $E(q,k) \ge 0 ( \le 0 ) $ in the electron-doped (hole-doped) region (, namely for the pole of the upper (lower) Hubbard branch ). 
%The fermionic Grassmann variable for electrons is described by $\phi$ and its conjugate $\phi^*$. 
The imaginary part of the self-energy ${\rm Im} \Sigma$ and the renormalization factor $Z$ is not considered here, because the singularities of the Mott transition are our main interest in this article while we assume that the singularities are not altered by ${\rm Im} \Sigma$ and $Z$~\cite{AssaadZ}. 
%$\Delta_c$ vanishes at the transition point and the Mott transition may be driven by the sign change in $\Delta_c$.

Now we take the hole picture for the hole doping side so that $E$ always takes $E\ge 0$ both in electron and hole doped regions.  Then, aside from the rigorous validity of the details of the above form for $E$, around the Mott gap edge, one can assume that the dispersion is expanded in terms of $k$ as 
\begin{equation}
E(q,k)=a(q)k^2 + b(q)k^4+..... 
\label{Disp2}
\end{equation}
We have shifted the chemical potential to cancel the gap $\Delta_c$.
Here, the gap edge is not necessarily isolated points in the momentum space, but may be a line or a surface, which may evolve to the Fermi surface in the Fermi liquid. The $k$-linear term does not exist because we have assumed that $k=0$ is the gap edge: The $k$-linear term violates the requirement $E\ge  0$ for small negative $k$.  The cubic term should also be vanishing in the region of our interest because it becomes relevant only when $a$ becomes sufficiently small, while then the cubic term also violates the initial assumption of $E(q,k)\ge 0$ for negative $k$.  The analyticity of the dispersion at small $k$ is our assumption.  This is actually plausible when the transfer energy for the distant pair of Wannier orbitals well converges to zero with increasing distance in the Hubbard-type models.  
 
The coefficients $a$ and $b$ are obtained as renormalized values after eliminating the higher order terms of the quasiparticle operator.
In the metallic side, the rigid band picture is not justified. However, it is still legitimate to consider the quasiparticle dispersion around the Fermi level and the coefficients $a$ and $b$ as effective quasiparticle coefficients obtained in the evolution process of the metallic phase. This means that $a$ and $b$ may depend on the distance from the Mott transition point, whereas they still behave continuously.  It should be noted that the variations of $a$ and $b$ can again be renormalized, which leads to $\zeta$ independent $a$ and $b$ near the transition point. 

If the gap edge is given by isolated points and the coefficients $a(q)$ at these gap edges approach nonzero positive constants on the verge of the transition, the dynamical exponent $z$ characterized by the single-particle dispersion is given by $z=2$ as in the generic transition to the band insulator~\cite{RMPX} with a finite effective mass of quasiparticles. We see below that this does not hold any more at the marginal quantum critical point.

From this quasiparticle description, the transition between metals and Mott insulators are
described by the change in the quasiparticle dispersions. One of our central statements in this paper is that the criticality of the transition has one to one correspondence with the singularity of the quasiparticle dispersions.  When the coefficient $a$ stays positive through the transition at finite number of isolated points of gap edge, the free energy has a similar singularity with the transition between the band insulator and metals, which is the continuous transition at zero temperature along the $T=0$ {\it boundary} in Fig~\ref{Fig1}(b).  The first-order transition may evolve only when $a$ becomes zero.  In the following, we clarify how the character of the transition and the single-particle dispersion are related each other.  

\subsection{Relation to Free Energy Form}
\label{III.D}
Here we relate the quasiparticle dynamics and the singularity of the free energy at the Mott transition in the quantum region.
When we take the path integral formalism with the imaginary time $\tau$, the singular part of the free energy density is formally written by using the quasiparticle dispersion $E$ as 
% ハバード模型のアクション
\begin{eqnarray}
F &=& -(T/N){\rm ln} Z \label{FreeEnergy}\\
Z &=& \int \prod_i {\cal D}\phi_i(\tau){\cal D}\phi_i^*(\tau){\rm e}^{-S/\hbar}, \\
S &=& \int_0^{\hbar/T}{\rm d}\tau \sum_i \phi_i^* \hbar \del_{\tau} \phi_i
+ \int_0^{\hbar/T}{\rm d}\tau H(\phi^*,\phi), 
\label{PART1}
\end{eqnarray}
where the effective hamiltonian $H$ generating the single particle excitation $E$ in Eq.~(\ref{Disp2}) is rewritten by using the Grassmann variables $\phi_i$ and $\phi_i^*$ at the site $i$.

The above quasiparticle form leads to the effective action at the chemical potential $\mu$  as
\begin{equation}
S=\sum_{i,n} \phi^*(q,k,\omega_n) (-i\omega_n - \mu + E(q,k))\phi(q,k,\omega_n)
\label{Action}
\end{equation}
with the Matsubara frequency $\omega_n$.

Let us first study the filling-control transition.  Although the expansion (\ref{1.4-2}) does not hold in the quantum region, the singular part of the free energy at zero temperature still has an expansion with respect to the doping concentration $X$. 
From the Matsubara-frequency and wavenumber dependent path integral form of the quantum dynamics leading from Eqs.(\ref{FreeEnergy}) to (\ref{Action}), we obtain for the singular part of the free energy density at the transition,
\begin{equation}
F =-X\mu-T\int_0^{\infty} dE D(E)\log(1+e^{-(E-\mu)/T})
\label{FreeEnergy2}
\end{equation}  
with $D(E)$ being the singular part of the density of states of quasiparticles. The Boltzmann constant is taken to be unity for our temperature scale.  The particle density measured from the 
insulating phase is given by 
\begin{equation}
X =\int_0^{\infty}dEf(E)D(E)
\label{ParticleDensity}
\end{equation}  
with the Fermi distribution function $f(E)\equiv 1/(e^{(E-\mu)/T}+1)$.

We first consider zero temperature and the case where the dispersion $E$ has minima at finite number of isolated points in the momentum space with the dispersion given by Eq.(\ref{Disp2}).
Then the particle density is given by 
\begin{equation}
X =A_dk_F^d,
\label{ParticleDensity2}
\end{equation} 
where $A_d$ is a dimensionality-dependent constant and $k_F$ is the Fermi wavenumber measured from the gap edge at the dispersion minima. Here we ignored the possible anisotropy of $k_F$ because it does not alter the essential part of the results below for the scaling properties.   

From the above relations, at $T=0$, we have
\begin{equation}
F =-X\mu+A_d[\frac{2a}{d+2}(\frac{X}{A_d})^{\frac{2}{d}+1}+\frac{4b}{d+4}(\frac{X}{A_d})^{\frac{4}{d}+1}+\cdots].
\label{FreeEnergy3}
\end{equation}  
We note that we are considering only the singular part of the free energy at the transition.

When the dispersion proportional to $a$ around these points $q_1$ in Eq.(\ref{Disp2}) is present, the quasiparticle picture predicts that the total free energy measured from the insulator has the lowest order term at $T=0$ as 
\begin{equation}
F+X\mu\propto aX^{(d+2)/d}.
\label{Disp3}
\end{equation}  
%This is obtained from the fact that the doping concentration is given from the integration over the phase space as 
%\begin{equation}
%X\propto \int_0^{k_F}k^{d-1}{\rm d}k,
%\label{Disp3-3}
%\end{equation}
%and the singular part of the free energy is similarly obtained from 
%\begin{equation}
%F+X\mu\propto\int_0^{k_F}k^{d-1}k^2{\rm d}k
%\label{Disp3-2}
%\end{equation}
%where we integrate over $k$ around $q_1$. Equation (\ref{Disp3}) is obtained by eliminating $k_F$ from Eqs.(\ref{Disp3-3}) and (\ref{Disp3-2}). The wavenumber $k_F$ means simply the wavenumber of the Fermi surface measured from the gap edge.  
From this free energy form, the charge susceptibility shows the scaling $\chi_c \equiv (\del^2 F/\del X^2)^{-1}\propto X^{1-2/d}$, which is the same as the transition to the band insulator. Then the first-order transition does not take place, because the charge susceptibility does not diverge for $d \ge 2$.
The metal insulator transition is well defined only at zero temperature.  This means that the transition occurs across the $T_c=0$ boundary illustrated in Fig.~\ref{Fig1}.

By starting from this continuous transition at zero temperature, the first-order transition can evolve in two fashions.
One possibility is the case where a large Fermi surface satisfying the Luttinger theorem appears immediately upon doping. In this case, one has to take that Fermi surface as the locus $E(q,k)-\mu=0$ and one gets
\begin{equation}
F+X\mu\propto X^{3}
\label{Disp3-4}
\end{equation}
by eliminating $k_F$ from
\begin{equation}
X\propto\int_0^{k_F}{\rm d}k
\label{Disp3-6}
\end{equation}
and 
\begin{equation}
F+X\mu\propto\int_0^{k_F}k^2{\rm d}k,
\label{Disp3-5}
\end{equation}
where we  perform the integrations in the region around the locus $E-\mu=0$ with the assumption $a>0$ everywhere.
In this case, one gets the charge susceptibility $\chi_c \equiv (\del^2 F/\del X^2)^{-1}\propto X^{-1}$, which means that the transition occurs at the {\it marginal quantum critical point}, where $T_c$ is still zero.  This is because $\chi_c$ diverges only at $X=0$ at zero temperature.  To realize the phase separation (namely, the first-order transition), we need a further additional degeneracy of the dispersion at the gap edge.   

The second possibility is the marginal quantum critical point emerging with the vanishing $a$ term at some isolated points $q_0$ of the gap edge. In this case, we have the lowest order term
\begin{equation}
F+X\mu\propto bX^{(d+z)/d},
\label{Disp4}
\end{equation}
with $z=4$, which yields 
\begin{equation}
\chi_c \propto X^{1-z/d}.
\label{Disp4-1}
\end{equation}
The exponent $z=4$ appears because we are left with the quartic term proportional to $b$ when the quadratic term proportional to $a$ vanishes.

\subsection{Electron differentiation}

Now it turns out that the two possible ways of realizing the first-order transitions eventually become merged to a unified picture, because even when we have the locus of $E-\mu=0$ with Eq.(\ref{Disp3-4}) being satisfied, a further flattening of the dispersion at the gap edge with vanishing $a$ is required to realize the first-order transition.  It is unlikely that such a flattening emerges uniformly on the locus $E=0$.  Instead, it generically occurs from particular points of the locus because of the initial anisotropy of the Fermi surface in the band structure and the anisotropic correlation effects as well.  Namely, the point with the smallest amplitude of $a$ becomes zero first as a special point of the $E-\mu=0$ surface.  Then a quartic dispersion appears at this special point of the $E-\mu=0$ surface when the system becomes marginal, which results in $z=4$.  

Therefore, the Mott criticality of the marginally quantum critical point is also characterized by an inevitable evolution of the electron differentiation, if the large Fermi surface is involved in the metallic side.  The singular differentiation generates a quartic dispersion at particular points of the Fermi surface coexisting with dispersive generic part.

\subsection{Comparisons with numerical and experimental results}

This large dynamical exponent $z=4$ was suggested in several independent numerical calculations for the filling-control transition of the Hubbard and $t$-$J$ models in two dimensions at $T=0$ ~\cite{RMP,Imada1995}.
These are the exponent estimated from the compressibility in the form (\ref{Disp4-1})~\cite{FurukawaImada0,FurukawaImada,FurukawaImada2,Kohno}, the Drude weight $D$ estimated from the form $D\propto X^{1+(z-2)/d}$~\cite{Tsunetsugu,Nakano}, single-particle dispersion~\cite{Assaad99} and the localization length $\xi\propto (\mu-\mu_c)^{-1/z}$ in the insulator side estimated from Green's function Eq.(\ref{Disp1}), where $\xi$ is obtained from the Fourier-transformed spatially-dependent Green's function $G(r,\omega=0)\sim \exp(-r/\xi)$~\cite{Assaad96}. 
These imply that the Hubbard and the $t$-$J$ models are located close to this marginal quantum critical point.

Although the quasiparticle picture does not hold, it has numerically been shown that the dynamical exponent indeed becomes $z=4$ at the marginal quantum critical point in the one-dimensional Hubbard model with next-nearest-neighbor transfers~\cite{25}. This suggests that the form (\ref{Disp4}) is universally valid irrespective of the applicability of the quasiparticle picture.  

It is insightful to compare experimental results obtained for the high-$T_c$ cuprates with the present picture of the electron differentiation. 
In the high-$T_c$ cuprates, flat dispersions are universally observed near $(\pi,0)$ and $(0,\pi)$ points in the angle-resolved photoemission experiments~\cite{7,8}.  
This flatness is beyond the conventional expectation obtained from the van Hove singularity, while it is a natural consequence, if the Mott critical point at $T=T_{c}$ is located at low temperatures. At and below $T_c$, vanishing quadratic dispersion, given by $a=0$ should emerge in a region of the expected Fermi surface.  

The distance from the marginal quantum critical point in the phase diagram in Fig.~\ref{Fig1} may depend on details of materials and models.
In addition, the actual $q_0$ positions responsible for the marginal quantum Mott criticality may also depend on materials and models.  For example, the change in relative amplitude of $t'$ to $t$ in the Hubbard model (\ref{1.3}) may change the location of $q_0$. This may even change the nature of the transition from the route across the $T_c=0$ boundary to the route through the first-order transition illustrated in Fig.~\ref{Fig1}. This change may indeed occur from one high-$T_c$ compounds to another~\cite{Andersen,Imada2004}.  Actually the singular points $q_0$ may deviate from $(\pi,0)$ and $(0,\pi)$ points for larger $t'$.  Width of the critical region may be influenced by the amplitude of $t'$ as well. 

For the moment, definite assignments are not possible, but we infer two possibilities for the cuprates.  One is that the system is indeed close to the marginal quantum critical point and the Mott criticality is controlled by the flat dispersion. Even when the Mott criticality is controlled by these flat points, the experimental observation of the flat part of the dispersion at the Fermi level may have some difficulty because the strong damping is inevitably accompanied.  The arc structure is observed in the angle-resolved photoemission experiments for the underdoped cuprates, which literally means that the Fermi surface around the flat-dispersion region is missing.  This implies that the experimental resolution might not allow the detection of the Fermi surface around the flat part, because of the strong damping, while this flat part may govern the criticality.

The other possibility is that the part of the flat dispersion is slightly away from the Mott gap edge. This is inferred from the fact that the flat-dispersion level in the hole-doped cuprates is slightly lower than the arc part around $(\pi/2,\pi/2)$, which is the dispersion minimum (in the hole picture) at the real Mott gap edge indicated by experiments~\cite{8,Ino} and by model calculations as well~\cite{AssaadZ}. In this case, with the lowering doping concentration, the system first shows the marginal quantum critical behavior reflecting the flat dispersion.  However, with further approaching the real critical point, it crossovers from the marginal quantum criticality characterized by $z=4$ to the ordinary class $z=2$, which eventually flows to the criticality for the $T_c=0$ boundary.  The real high-$T_c$ cuprates appears to have a variety between these two possibilities depending on the compounds.  We will discuss consequences of the latter case further in Sec. \ref{long-rangeCoulomb}.

Aside from these details and uncertainties, the overall structure of very different evolutions of the Fermi surface depending on the momenta is completely consistent with the picture that the transition metal compounds including the cuprates show electron differentiation arising from the proximity of the marginal quantum Mott criticality. Electron differentiation should become prominent when the system becomes closer to the marginal quantum critical point, while the electrons are more or less uniform along the $T_c=0$ boundary.

\subsection{Breakdown of Ginzburg-Landau-Wilson scheme}

From the results in Secs.~\ref{III.C} and \ref{III.D}, the free energy near the marginal quantum critical point is generally expressed as
\begin{equation}
F=-\mu X+aX^{(d+2)/d}+ b X^{(d+4)/d},
\label{Disp5}
\end{equation}
where $a$ and $b$ have absorbed numerical constants in Eq.(\ref{FreeEnergy3}) as well as effects of the renormalization factor $Z$ being less than unity, and the constants have properly been rescaled.

Even in the case of the bandwidth-control transition, when one can regard the closing of the gap by hole doping around a point $q_{0h}$ and simultaneous particle doping around $q_{0p}$ with the constraint of keeping the electron density $n=1$, the above relation Eq.(\ref{Disp5}) may be replaced with $D$ as 
\begin{equation}
F= UD+aD^{(d+2)/d}+ b D^{(d+4)/d},
\label{Disp6}
\end{equation}
because the ``doublon" and ``holon" concentrations are nothing but the above self-doping concentration of particles and holes.  

%フィリング制御で引力の原因は？コメンシュラビリティ？反強磁性ゆらぎ？

Now instead of Eq.(\ref{1.4-2}), the free energy at zero temperature is expanded by $\zeta$ and obtained after rescaling of the parameters as 
\begin{equation}
F= -\mu_{\zeta}\zeta + a\zeta^{(d+2)/d} +b\zeta^{(d+4)/d}+c\zeta^{(d+6)/d}\cdots
\label{QuantumFreeEnergy}
\end{equation}
 with the constraint $\zeta\ge0$.
It should be noted that this expansion of the free energy in terms of $\zeta$ is obtained from
the path integral form with the spatial as well as imaginary time dependence explicitly taken into account.
This form of the free energy clearly violates the Ginzburg-Landau-Wilson scheme because the form of the free energy itself has $d$ dependent nonanalytic expansion. 

%Instead of the classical GL expansion;

The metal-insulator transition across the $T_c=0$ boundary with $a>0$ is driven by the $\mu_{\zeta}$ term in Eq.(\ref{QuantumFreeEnergy}).  When $\mu_{\zeta}$ is negative, the free-energy minimum exists at $\zeta=0$, which corresponds to an insulator. Whereas the metallic phase is represented by the minimum at a nonzero positive $\zeta$, which is realized by a positive $\mu_{\zeta}$.  The $T_c=0$ boundary is determined from $\mu_{\zeta}=0$.
The criticality of this $T_c=0$ boundary is given for the order parameter as  
\begin{equation}
\langle \zeta\rangle \propto|\mu_{\zeta}|^{\beta}.
\end{equation}  
In the ``mean-field approximation", $\beta$ is obtained from the spatially uniform derivative $\partial F /\partial \zeta =0$ as
\begin{equation} 
\beta=d/2.
\label{beta0}
\end{equation}
The susceptibility is given by 
\begin{equation}
\chi_{\zeta}=[\frac{d^2F}{d\zeta^2}]^{-1}\sim [\frac{2(d+2)}{d^2}a\zeta^{2/d-1}]^{-1},
\end{equation}
leading to the susceptibility exponent defined by
\begin{equation}
\chi_{\zeta}\propto|\mu_{\zeta}|^{-\gamma}
\label{chiscale}
\end{equation}
in the metallic side $\mu_{\zeta}<0$ as
\begin{equation}
\gamma=1-d/2.  
\label{gamma0}
\end{equation}
At $\mu=0$, we obtain another exponent from 
\begin{equation}
\chi_{\zeta}\sim a^{-1}\zeta^{1-\delta}
\end{equation}
 with 
\begin{equation}
\delta =2/d
\label{delta0}
\end{equation}
in the metallic side.  These are all ``mean-field exponents", although we will show these exponents are indeed correct.
In the Ginzburg-Landau mean-field theory, the critical exponents do not depend on the dimensionality.  However, this $d$-dependent form of the free energy leads to the $d$ dependent exponents even in the mean-field treatment.  

When $a$ becomes zero, the critical point $\mu_{\zeta}=0$ becomes marginal and the first-order transition evolves if $a$ becomes negative.  This point with $\mu_{\zeta}=a=0$ at $T=0$ is nothing but the {\it marginal quantum critical point}.  Namely, the marginal quantum critical point at $T=0$ may be reached at a control parameter $g=g_c$, for $a=a_0(g-g_c(T))$, $a_0>0$ and $b>0$.

One might argue that the marginal quantum critical point looks similar to the conventional tricritical point~\cite{Tricritical} because the continuous transition converts to the first-order transition at $T=0$ at this point.  However, it is qualitatively different because the Mott transition contains only the insulator and metal phases and no additional competitions a priori exist. 

Similarly to the transition across the $T_c=0$ boundary, the critical exponents of the marginal quantum critical point have $d$ dependent forms even in the mean-field treatment.  The exponent $\beta$ defined by the order parameter at $g<g_c$ and $\mu_{\zeta}=0$ as 
\begin{equation}
\langle \zeta\rangle \propto|g-g_c|^{\beta}
\end{equation}  
is given by 
\begin{equation} 
\beta=d/2.
\label{beta}
\end{equation}
Near $g=g_c$, $\chi_{\zeta}$ is expressed as 
\begin{equation}
\chi_{\zeta}=[\frac{d^2F}{d\zeta^2}]^{-1}\sim [\frac{2(d+2)}{d^2}a_0(g-g_c)\zeta^{2/d-1}+\frac{4(d+4)}{d^2}b\zeta^{4/d-1}+\frac{6(d+6)}{d^2}c\zeta^{6/d-1}]^{-1}.
\end{equation}
Then 
\begin{equation}
\chi_{\zeta}\propto|g-g_c|^{-\gamma}
\label{chiscale}
\end{equation}
holds in the metallic side $g<g_c$, yielding the ``mean-field" exponent 
\begin{equation}
\gamma=2-d/2.  
\label{gamma}
\end{equation}
At $g=g_c$, \begin{equation}
\chi_{\zeta}\sim b^{-1}\zeta^{1-\delta}
\end{equation}
 with 
\begin{equation}
\delta =4/d
\label{delta}
\end{equation}
is obtained. We note that the divergence of the susceptibility is stronger at the marginal quantum critical point than along the $T_c=0$ boundary.  It is also stronger for lower spatial dimensions.
For example, $\gamma$ is 1 and 1/2 in two and three dimensions, respectively, for the marginal quantum criticality. On the contrary,$\gamma$ is not positive in two and three dimensions along the $T_c=0$ boundary. 

Even though the Ginzburg-Landau-Wilson scheme breaks down, the scaling relations $\beta\delta=\gamma+\beta$ and $\alpha+2\beta+\gamma=2$ are satisfied both for the $T_c=0$ boundary and the marginal quantum critical point.  Along the $T_c=0$ boundary we obtain $\alpha=1-d/2$ while $\alpha=-d/2$ for the marginal quantum criticality.  They are consistent with the hyperscaling law $2-\alpha=(d+z_t)\nu$ with $\nu=1/2$ and $z_t=4$ for the marginal quantum criticality and $z_t=2$ along the $T_c=0$ boundary.
Here, $z_t$ should be the dynamical exponent of this Mott transition, namely the dynamical exponent for the density or doublon density correlations, which turns out to coincide with the dynamical exponent for the single particle excitations, $z$ determined from the quasiparticle dispersion.    
   
We can also confirm that the hyperscaling relation is satisfied in the following way:
The scaling relation and the exponents are derived from the scaling form of the free energy, 
\begin{equation}
F(a,\mu_{\zeta})=\xi^{-d-z_t}f(a\xi^{y_g},\mu_{\zeta}\xi^{y_{\mu}})
\label{ScalingFunction}
\end{equation}
with a scaling function $f$ and the correlation length 
\begin{equation}
\xi\propto (g-g_c)^{-1/2}\propto \zeta^{-1/d},
\label{xiscale}
\end{equation}
which implies $y_g=2$.
Here the crossover exponent $y_{\mu}=4$ ($y_{\mu}=2$) is derived from the dynamical exponent of the density fluctuations given by $z_t=4$ ($z_t=2$) for the marginal quantum criticality ($T_c=0$ boundary), respectively. The hyperscaling relation holds because this scaling form (\ref{ScalingFunction}) is satisfied. In fact this scaling form is derived from the single length scale $\xi$ which diverges at the transition point. This correlation length is indeed proportional to the mean carrier distance $\propto X^{-1/d}$ in the filling-control transition. This is obviously the single length scale which diverges at the transition.  Satisfaction of the hyperscaling form (\ref{QuantumFreeEnergy}) also clearly shows that this criticality is the consequence of the spatio-temporal quantum dynamics of two-particle excitations in the path integral form, where $d$ and $z_t$ represent spatial and imaginary-time fluctuations, respectively.

\subsection{Two-dimensional case}
In one and two dimensions, the powers of expansions in Eq.(\ref{QuantumFreeEnergy}) stay at integers (for example, in 1D, $(d+2)/d$ reduces to 3 and in 2D, $(d+2)/d$ reduces to 2).  
In two dimensions, the free energy is reduced to 
\begin{equation}
F=-\mu_{\zeta}\zeta+a_0(g-g_c)\zeta^2+b\zeta^3+c\zeta^4.
\end{equation}
This again does not belong to the conventional scheme of the Ginzburg-Landau-Wilson formalism,
because the odd order term (the cubic term here) is not allowed in the conventional Landau expansion from the constraint of the symmetry around the critical point.  This has some resemblance to the breakdown of the Ginzburg-Landau-Wilson scheme at the Lifshitz point of the structural transition although the physics contained here is quite different.  Here the asymmetry is allowed because the part of negative $\zeta$ does not exist.  This can be easily understood in the analogy to the trivial metal to the band-insulator transition, where the carrier density cannot be negative either. Even in the transition between metals and band insulators in the noninteracting systems, the free energy has a similar form to Eq.(\ref{QuantumFreeEnergy}) and the Ginzburg-Landau-Wilson scheme does not hold. 

In the Mott transition in two dimensions, the susceptibility is given by
\begin{equation}
\chi_{\zeta}=\left(\frac{d^2F}{d\zeta^2}\right)^{-1}\sim \frac{1}{2a_0(g-g_c)+6b\zeta+12c\zeta^2}.
%\nonumber
\label{2}
\end{equation}
Then, $\gamma=1, \beta=1$ and $\delta=2$ hold~\cite{Imada2004,Imada2005}. 
Remarkably, this agrees with recent experimental results on a $\kappa$-ET compound, $\kappa$-(ET)$_2$Cu[N(CN)$_2$]Cl by Kagawa, Miyagawa and Kanoda~\cite{Kanoda,Kanoda3}.   The exponents of the finite-temperature critical point estimated by the conductance are indeed consistent with these values for $\beta, \gamma$ and $\delta$  within the experimental accuracy. We note that this compound has a structure of highly two-dimensional anisotropy.  For the moment, it is not well clarified how the crossover to the three dimensionality arising from weak three-dimensional coupling should appear experimentally. 
We will discuss below that these unusual exponents are also obtained practically even at finite temperatures, which is relevant in the realistic experimental condition of the finite-temperature Mott transition.
%In addition to the temperature, doping and pressure dependences, 
%the exponent can also be measured by the electric-field induced transition 
%through the nonlinear conductivity.
  
\subsection{Validity of the mean-field theory}
\label{III.I}
Very close to the marginal quantum critical point, the present mean-field exponent only marginally breaks down, because the Ginzburg criterion~\cite{Goldenfeld} $d+z_t\ge(2\beta+\gamma)/\nu$  with $\nu=1/2$ being the correlation length exponent indicates that the system is always at the upper critical dimension irrespective of the dimensionality, because the equality $d+z_t=(2\beta+\gamma)/\nu$  always holds. Here, $\nu=1/2$ is a direct consequence of $y_g=2$. 
The fluctuation beyond the mean-field theory becomes irrelevant above the upper critical dimension.   Although logarithmic corrections may exist, the mean-field description is thus basically correct at any dimension in this case.  
This explains why the ``mean-field exponents" are observed in the $\kappa$-ET compound.
Below we restrict our analysis to the mean-field study because the primary exponents are correct. 
Although the mean-field exponents are correct, the hyperscaling also holds at any dimension.  This peculiar compatibility is explained by the fact that the system is always at the upper critical dimension at any dimension.
Detailed analysis of corrections based on the renormalization group study will be reported elsewhere.
The same argument for the validity of the mean-field exponents is applied for the $T_c=0$ boundary.

Although the diverging density fluctuation is an inevitable consequence of the Mott critical point for the marginal quantum criticality and the classical Ising criticality, it is highly nontrivial effect from the viewpoint of the weak coupling picture. In fact, naive perturbation expansions result in suppressions of the density fluctuations when the Mott transition is approached and the available perturbative treatment fails in reproducing this criticality. The one-loop calculation does not account for the Mott criticality, which in principle has to be explained in the real part of the self-energy to be calculated from the higher order loops, while it is so far an open issue to be derived in the future.   In this sense, even the mean-field theory is not a straightforward framework in contrast with most of the mean-field theories as those in magnetic transitions.

\subsection{Finite temperature effect}
The parameter $g$ also has temperature dependence, because the quasiparticle dispersion in general has a temperature dependence. As we have clarified above, $g$ is determined from the quasiparticle dispersions.  Therefore $g-g_c$ can be replaced by $T-T_c$, where the critical temperature becomes nonzero. This is the dominant finite-temperature effect at low temperatures.
However, we have quite independent origin of the finite-temperature effects originating from the entropy term $T\zeta \ln \zeta$ in addition to the above temperature dependence in the free energy $F$.  
This generates an essentially singular contribution for $\mu_{\zeta}<0$, because the free energy has the extremum at $\zeta_0 \propto \exp[\mu_{\zeta}/T]$.  This is an exponentially small contribution at low temperatures for $\mu_{\zeta}<0$. Therefore, it does not contribute to the present scaling behavior near $T=0$ in the power of $T$. 
At high temperatures, however, without the Fermi degeneracy, the expansion around the extremum reproduces the regular Ginzburg-Landau form  
\begin{equation}
F_T=\frac{A_0(T-T_c)\zeta^2}{2}+\frac{B\zeta^4}{4}-\mu_{\zeta}\zeta,
\nonumber
\end{equation}
when we redefine as $\zeta-\zeta_0 \rightarrow \zeta$. 

When $T$ is nonzero, strictly speaking, the Ising universality class may appear in the region extremely close to the critical point even at low temperatures. However, this real critical region becomes exponentially narrow with decreasing temperatures and the quantum criticality governs out of this region. This crossover of the criticality is due to the essentially singular contribution of the entropy term.

The results obtained for V$_2$O$_3$~\cite{Limelette} indicates that $T_c$ is above 400 K and is high enough so that the Ising classical universality is well observed. On the other hand, $T_c$ for $\kappa$-ET$_2$Cu[N(CN)$_2$]Cl is below 40K and low enough~\cite{Kanoda}, so that the quantum region dominates in the experimental results.  All of these are consistent with what were observed. More quantitative analyses are left for future studies.

\subsection{Effects of long-range part of Coulomb interaction for filling-control transitions}
\label{long-rangeCoulomb}

The role of the long-range part of the Coulomb interaction between electrons is sometimes controversial.  In this subsection, we clarify how the long-range interaction modifies the conclusions obtained for the models only with the short-range interaction.  
Because the phase separation occurs for models with only the short-range force as the first-order transition in the classical region, we restrict ourselves to the case of the Ising classical universality in this subsection, Sec.\ref{long-rangeCoulomb}.  Because of the long-range part and the resultant electrostatic condition, the real phase separation into two different electron densities is not allowed in the filling-control Mott transition.  Reflecting the electrostatic condition, the diverging charge susceptibility at strict zero wavenumber is eventually suppressed for the filling control transitions. However, it still causes critical fluctuations at nonzero and small wavenumbers.

The Poisson equations for the external test charge $\rho_{ext}$ and the induced charge $\rho_{ind}$ are given by 
\begin{eqnarray}
\nabla^2\phi_{ext}& =& -\frac{\rho_{ext}}{\epsilon}, \nonumber \\
\nabla^2\phi_{ind} &=& -\frac{\rho_{ind}}{\epsilon},
\label{1.5} 
\end{eqnarray} 
where $\phi_{ext}$ is the electrostatic potential generated by the external test charge, while the induced charge generates additional potential $\phi_{ind}$. The dielectric constant is $\epsilon$. When the external test charge $\rho_{ext}$ is at the origin with the unit charge, in metals, 
the Thomas-Fermi screening occurs as $\rho_{ind}=-\alpha \phi_{tot}$ for weakly $k$-dependent term with the definition of the total electrostatic potential $\phi_{tot}=\phi_{ext}+\phi_{ind}$.
In fact, by the Thomas-Fermi screening, $\alpha$ is estimated as
$\alpha=2meq_F/(\pi\hbar^2)$, where $m,e$ and $q_F$ are the electron effective mass, charge and the Fermi wavenumber, respectively.  Here we have assumed isotropic sphere of the Fermi surface for simplicity. 
Since the external test charge at the origin leads to a wavenumber independent form $\rho_{ext}(q)=1$ in the momentum representation, we obtain  
\begin{equation}
\phi_{tot}=\frac{1}{\alpha+\epsilon q^2}
\label{1.6} 
\end{equation} 
and the resultant induced charge as
\begin{equation}
\rho_{ind}=\frac{-\alpha}{\alpha+\epsilon q^2},
\label{1.7} 
\end{equation} 
which indicates the Yukawa-type normal screening, $\rho_{ind}(r)=\alpha\exp[-2\sqrt{\pi e\alpha} r]/r$ in the real space.
When one considers the susceptibility as the linear response induced by the imposed external charge as the perturbation, the susceptibility $\chi$ defined by $\rho_{ind}=-\chi\rho_{ext}$ is now given by 
\begin{equation}
\chi=\frac{\alpha}{\alpha+\epsilon q^2}.
\label{1.8} 
\end{equation} 
Whereas, if the Coulomb potential is regarded as the perturbation, the linear response defined by
$\rho_{ind}=-\chi_c\phi_{ext}$ is given by 
\begin{equation}
\chi_c=\frac{\epsilon \alpha q^2}{\alpha+\epsilon q^2},
\label{1.9} 
\end{equation} 
which shows that the charge susceptibility is suppressed near $q=0$ because of the long-range Coulomb force.
With this screening, the free energy in the classical region is modified to
\begin{equation}
F(r) = \int dr' \frac{1}{2}\chi_c^{-1}(r-r')X(r')^2+\frac{1}{4}bX(r)^4-e\phi X,
\label{1.10} 
\end{equation} 
where the Fourier transformed susceptibility is given by 
\begin{eqnarray}
\chi_c^{-1}(q)&=&\chi_{c{\rm H}}^{-1}(q)+\frac{1}{\epsilon q^2}, \label{1.11-1} \\
\chi_{c{\rm H}}^{-1}(q)&=&\frac{1}{\alpha}+c(\alpha)q^2.
\label{1.11} 
\end{eqnarray} 
Here, $\chi_{c{\rm H}}$ is the charge susceptibility for the Hubbard model without the long-range Coulomb part.
The second term in the right hand side of Eq.(\ref{1.11-1}) represents the long-range Coulomb interaction. 
The scaling of $\alpha$ near the critical point is determined from the effective mass $m$ and is given as $\alpha \propto \xi^{4-d}$ and the function $c(\alpha)$ scales as $c(\alpha)\propto \xi^{2-d}$. The charge susceptibility is enhanced except for extremely small wavenumber.  

In the realistic condition, the minimum of Eq.(\ref{1.10}) appears at $q_{min}\sim 1/(\epsilon c)^{1/4}$.  In the cuprates, this is roughly estimated to be $q_{min}\sim 0.1\pi$, since $\epsilon$ is the order of 10-100, and $c\sim 10$ as we see below.     Namely, the charge susceptibility becomes strongly enhanced in the region of small wavenumbers, which are in the order of one tenth of the Brillouin-zone size, $2\pi$. Even around the marginal quantum Mott criticality, this induces a ``softening" of the charge response with poor screening in the nm length scale, which causes strong dynamical fluctuations of electron density at these small wavenumbers. This provides us with mechanisms of various unusual properties for metals near the Mott insulator. If $F$ becomes negative at $q_{min}$, an instability toward the charge ordering occurs. 

Even when $T$ is below $T_c$ in the Hubbard-type models, where only the short-range force is considered, the phase separation dynamics toward the $q=0$ mode in reality freezes at a stage of a finite $q$ constrained from the long-range repulsion for the filling-control transition.   

It should be noted that in the bandwidth-control transition, the divergence at $q=0$ indeed occurs. This is because the electrostatic condition is not violated for the diverging doublon susceptibility.

We now realize that the larger dielectric constant induces the larger enhancement of the charge susceptibility as obtained by substituting $q_{min}$ into Eq. (\ref{1.11-1}). Actually, if good metallic carriers independently exist in addition to the strongly correlated electrons which yields the doped Mott insulator, such  good metallic carriers efficiently screen the long-range Coulomb interaction of the correlated electrons.  This allows more possibility for the part of the correlated electrons to approach the marginal quantum critical point.  In fact, when the good metallic carriers perfectly screen the motion of the correlated electrons, it corresponds to the limit $\epsilon \rightarrow \infty$ in the above argument and the diverging density susceptibility for the correlated part of electrons can indeed occur at $q=0$, because the density fluctuations are completely compensated by the good-metallic carriers except for the onsite interaction.    This may occur in a two-band system, where one band has large bandwidth supplying good metallic carriers, and the electrons on the other band are strongly correlated near the Mott insulating phase.  Although it is not a simple two-band system, the situation in the hole-doped cuprate superconductors is in a sense ideal from this viewpoint, because upon the carrier doping, ``good metallic carrier" first appears around the $(\pi/2,\pi/2)$ region of the Fermi surface, while the carriers near $(\pi,0)$ region with strong correlation effects become doped in the presence of these itinerant carriers.  Then the instability for the phase separation or the inhomogeneity actually occurs as the inhomogeneity of the ``$(\pi,0)$ carriers" compensated by the "$(\pi/2,\pi/2)$ carriers". To realize this situation, it appears to be important to recognize that the cuprates are located close to the marginal quantum critical point, but strictly speaking located slightly in the side of the $T_c=0$ boundary.

Even when the charge ordering or the phase separation does not occur, the spatial inhomogeneity is easily driven by impurity potential or lattice distortions, because of the underlying enhanced density susceptibility. This may be relevant as the mechanism of the structure observed by scanning tunnel microscopy (STM) in the cuprates and manganites~\cite{9,10,11,12,13}. The spatial inhomogeneity in the long length scale (typically at 1 to 10 nm scale) observed experimentally cannot be explained by the naive Thomas-Fermi screening length $\lambda_{\rm TF}$, since nominally $\lambda_{\rm TF}$ is the order of $1/q_F\sim 0.3 $nm.  The inhomogeneous structrure may also be enhanced by the experimental condition probed at the surface.  The present results obtained from the Mott criticality has a tight connection to the approach from dynamical stripe fluctuations~\cite{20,21}, while the importance of the underlying Mott criticality has not been recognized in the literature. 

\section{Non-Fermi-Liquid Properties}
\subsection{Mode coupling theory}
Now we discuss the consequences of the enhanced density susceptibility near the marginal quantum critical point.  Within the mean-field theory, we assume the dominant part of the susceptibility in the vicinity of a small and nonzero momentum $Q$ and around the zero frequency as
\begin{equation}
\chi_{\zeta}(q,\omega)=\frac{\Gamma^{-1}}{-i\omega+D_s(K^2+(q-Q)^2+\cdots)},
\label{3}
\end{equation}
where $D_s$ is the diffusion constant of the density fluctuations, $K$ is the distance from the marginal quantum critical point, and $\Gamma$ is a constant. Near the critical point, they follow the critical scaling as $\Gamma^{-1}\propto\xi^{-d}, K\propto\xi^{-1}$ and $D_s\propto \xi^{-2}$, which reproduces the above scaling Eq.(\ref{chiscale}) with Eq.(\ref{xiscale}) at $\omega=0$.  This again satisfies the dynamical exponent $z_t=4$ because $\omega$ scales as $D_sK^2\propto \xi^{-4}$, while the scaling of $K$ reproduces $\nu=1/2$. Since it is at the upper critical dimension, this Ornstein-Zernike-type form is justified.
We note that the enhancement is much stronger for $d=2$ than $d=3$. 

It should be noted that the characteristic energy scale of this charge fluctuation is much larger than that of spin and orbital fluctuations, because $c(\alpha)q^2$ should typically have the energy scale of the Mott gap at the boundary of the Brillouin zone. 
This is clear because at the zone boundary, the density fluctuation requires generation of the spatially alternating doublons and holons. Then $D_s$ has the energy scale comparable to the Mott gap.   For the filling control transition, it may have an energy scale as large as several eV as in the case of the cuprates ($\sim 2$ eV), while for the bandwidth-control transition, this energy scale is not necessarily large.
The large energy scale of these Mott fluctuations explains many puzzling properties of metals near the Mott insulator as we will discuss below.

Since the dominant fluctuation occurs at small but nonzero wavenumbers, the conservation law as in the density conservation at $Q=0$ does not exist and the dynamical exponent stays at $z_t=4$ for the filling-control transition.  For the bandwidth-control transition, the real divergence at $q=0$ again does not involve the conservation law because the doublon density does not commute with the Hamiltonian and is not conserved.  Therefore, the dynamical exponent again stays at four and the form Eq.(\ref{3}) is justified.  This is in contrast with the ferromagnetic fluctuation at $Q=0$ for the spin fluctuations, where $i\omega$ in Eq. (\ref{3}) is replaced with $i\omega/\Xi$ with $\Xi \propto q$ so that the dynamical exponent increases by one~\cite{28}.  

Now we formulate a mode coupling scheme for the electron-density and doublon-density fluctuations originating from the Mott criticality around small $Q$. 
The inverse of the static susceptibility is obtained from Eq.(\ref{2}) as 
\begin{equation}
\chi_{\zeta}(Q,0)^{-1} =2a_0(T-T_c)+6b\langle\zeta_Q\rangle+12c\sum_q\langle\zeta_q^2\rangle,
\label{4}
\end{equation}
where we introduced the average of the density fluctuation as $\langle\zeta_q^2\rangle$. This is given from the fluctuation dissipation theorem as 
\begin{equation}
\langle\zeta_q^2\rangle=\frac{1}{2\pi}\int^{\infty}_{-\infty}d\omega \coth(\omega/2T){\rm Im}\chi_{\zeta}(q,\omega).
\label{5}
\end{equation}

\begin{figure}
%$$ \psboxscaled{360}{imadaFig2n.eps} $$
\begin{center}
\includegraphics[width=8cm]{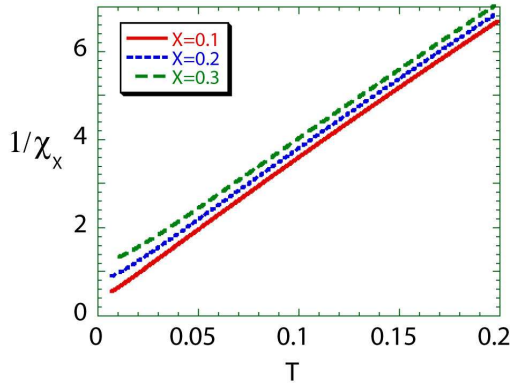}
\end{center}
\caption{\label{Fig.2}
Temperature dependence of the inverse of the density susceptibility peak.  A ``Curie-Weiss" temperature dependence is observed in an extended region for several choices of doping $X=0.1,0.2$ and 0.3. This linear temperature dependence causes non-Fermi-liquid properties.}
%\label{Fig.2}
\end{figure}
In the inverse of the bare susceptibility, the mass term is renormalized by the mode-coupling term 
proportional to  $\sum_q\langle\zeta_q^2\rangle$ in Eq.(\ref{4}). By renormalizing the zero temperature value, $T_c$ is renormalized to 
\begin{equation}
T'_c=T_c-\frac{6c}{a_0}\sum_q\langle \zeta_q^2\rangle (T=0),
\end{equation}
where 
\begin{equation}
\langle \zeta_q^2\rangle (T=0) \equiv \frac{1}{\pi}\int^{\infty}_{0}d\omega {\rm Im}\chi_{\zeta}(q,\omega)
\end{equation}
is the zero-temperature fluctuation.  Then the renormalized inverse susceptibility is given by
\begin{equation}
\chi_{\zeta}(Q,0)^{-1} =2a_0(T-T'_c)+6b\langle\zeta_Q\rangle+12c\sum_q\delta \langle\zeta_q^2\rangle
\label{4-1}
\end{equation}
with the definition for the finite-temperature correction;
\begin{equation}
\delta \langle\zeta_q^2\rangle=\frac{2}{\pi}\int^{\infty}_{0}d\omega \frac{1}{\exp(\omega/T)-1}{\rm Im}\chi_{\zeta}(q,\omega).
\label{5-1}
\end{equation}

We can solve Eqs.(\ref{3}), (\ref{4-1}) and (\ref{5-1}) selfconsistently, where we take into account the Gaussian fluctuations of the density through the mode coupling. In the mode-coupling scheme, the critical exponents stay at the mean-field values, which is justified from the above arguments in Sec. \ref{III.I}. However, effects of the fluctuations are taken into account in a selfconsistent fashion. 
A similar phenomenological theory has been formulated by Moriya for spin fluctuations~\cite{28}. 
 
\subsection{Perturbative self-energy}
In the following, we employ a perturbative treatment to understand the non-Fermi-liquid behavior as well as superconducting instability.  The perturbative scheme has a limited applicability for these highly nonperturbative phenomena of the density fluctuations.  However, when the Mott criticality is properly taken into account through the phenomenological treatment of the density fluctuations, we expect that the essence can be captured even when we employ a perturbation theory for the other part of calculations.

In the actual calculations for the realistic choice of parameters, a reasonable set of parameters can be derived in the following:  We choose the parameter values of a filling-control transition for a two-dimensional system appropriate for the copper oxide superconductors inferred from the frequency dependence of the optical conductivity~\cite{29,30}, characteristic size of the observed inhomogeneity~\cite{9,10} and the doping dependence of the density susceptibility in numerical\cite{FurukawaImada} and experimental results, which suggest $a_0\sim0, b\sim0.7, c\sim100,\Gamma^{-1}\sim3X$ and $D_s\sim30X$ by taking the energy unit $t (\sim0.4$eV) and the lattice constant as the length unit.  The characteristic wavenumber and energy of the density fluctuations are roughly $\pi/10$ and 0.5-1eV, respectively, which determines $b$ and $D_s$, respectively. In fact, the dielectric function $\epsilon(q=0,\omega)$ obtained from the optical conductivity provides us with ${\rm Im}\chi(q=0,\omega)={\rm Im}[1/\epsilon (q=0,\omega)]-1$.  The obtained results for ${\rm Im} \chi$ deduced from the experimental data for the optical conductivity have a prominent peak structure around 0.5-1 eV~\cite{30}, which indicates the characteristic energy scale of the density fluctuations. After considering the screening effect by the long-range Coulomb part, this suggests $D_s$ has the order of 1eV. The uncertainty of the parameters remains because of the lack of accurate experimental probes to estimate the frequency and wavenumber dependences of dynamical density fluctuations.  Basically, all the results presented here do not depend on $Q$ within the choice $0<Q<0.2\pi$.

As an example, we consider the carrier doping in the Hubbard model with the dispersion of the square lattice $E(q)=-2t(\cos q_x+\cos q_y)$ with additional input of the density fluctuations given by Eq.(\ref{3}).
Through the mode coupling, the solution of the selfconsistent equations (\ref{3}),(\ref{4-1}) and (\ref{5-1}) shows that the Curie-Weiss type behavior $\chi_{\zeta}\sim(T+\Theta)^{-1}$  holds in an extended temperature region with small Weiss temperature $\Theta$ near the marginal quantum critical point. Figure \ref{Fig.2} shows calculated results of such Curie-Weiss bahaviors in an extended temperature region. The criticality stays at the mean-field form, while the Curie-Weiss form is retained over very large temperature region with renormalized values of coefficients.  
Even when $a_0=0$ is employed, we obtain the linear temperature dependence of $\chi_{\zeta}^{-1}$, namely the Curie-Weiss form in a wide temperature region because of the linear temperature dependence of $\delta \langle\zeta_q^2\rangle$.  

Now we calculate the electron self-energy.
The electron self-energy in the perturbation expansion up to the second order of the interaction for the filling control is given as
\begin{equation}
\Sigma(q,\omega_n)=\frac{TU^2}{2N}\sum_{k,n}G(k,i\omega_n)\chi_{X}(q-k,i(\omega_n-\omega_m)).\label{6}
\end{equation}
Here the imaginary part of the self-energy ${\rm Im}\Sigma$ is governed by the Curie-Weiss behavior of $\chi_X$ through Eq.(\ref{6}).  Because of the linear temperature dependence in $\chi_{\zeta}^{-1}$ in an extended temperature region, we obtain the linear temperature dependence also for ${\rm Im}\Sigma$ in a wide temperature region.  In the marginal-critical region with $\chi_X\propto X^{-1}$, the standard Fermi-liquid behavior ${\rm Im} \Sigma \propto T^2$ is replaced with the non-Fermi-liquid form ${\rm Im} \Sigma \propto T$, which may cause various unusual properties.  The resistivity in two-dimensional systems becomes nearly proportional to $T$ as $\rho\propto {\rm Im} \Sigma \propto T$ in contrast to the standard Fermi-liquid scaling $\rho\propto {\rm Im} \Sigma \propto T^2$ .  

A long-standing puzzle in the doped Mott insulators is widely observed long-tail structures in the optical conductivity extending up to the order of 1 eV in various transition metal oxides and organic conductors~\cite{RMP,29,30,Lee}.  The tail has a structure of power law decay in the optical conductivity as $\sigma(\omega)\sim\omega^{-p}$ with $p$ ranging between 0.3 and 1.  Origin of such long tail structure has to be attributed to fluctuations in the energy scale of 1 eV and cannot be accounted for by the spin and orbital fluctuations, since they have much lower energy scale typically less than 0.1 eV.  We note that the density fluctuation mechanism examined in this article naturally accounts for such fluctuations at large energy scale.  
%More detailed study will be reported elsewhere.

\section{Superconductivity emerging from Mott criticality}
\subsection{Pairing originating from the marginal quantum Mott criticality}
It is widely recognized that the origin of the high-temperature superconductivity in the copper oxides~\cite{Bednorz} has to be explained by considering the strong electron correlation effects, although the mechanism is still puzzling and not definitely figured out.  After the discovery, various aspects of magnetic mechanisms were extensively examined.  From the weak coupling picture~\cite{Moriya,Pines}, the spin fluctuation theories were considered, where strong antiferromagnetic fluctuations were assumed to mediate the Cooper pairing.  Then the origin of the high-$T_c$ superconductivity was assumed to arise from the criticality of the antiferromagnetic quantum critical point.  In the strong-coupling expansion represented by the $t$-$J$ model~\cite{AndersonBaskaran}, it was claimed that the pairing is basically through the singlet formation stabilized by the superexchange term proportional to $J$ in the $t$-$J$ model.  In both of the approaches, the mechanism of the Cooper pairing is more or less the same and they are categorized as the magnetic mechanism.

On the other hand, from the initial stage of the studies on the cuprate superconductors, it has been well recognized that the superconductivity occurs in the region of the doped Mott insulator near the Mott transition~\cite{Anderson}.  However, since the criticality of the Mott transition was not well identified until recently, the role of the Mott criticality for the mechanism of the superconductivity was not well appreciated.  In fact, as we already clarified in the previous sections, the Mott transition itself is a transition driven by the order parameter of the electron (or doublon-holon) density, and has nothing to do with the magnetic degrees of freedom by itself.  Although the antiferromagnetic fluctuations occur at $(\pi,\pi)$ in the magnetic Brillouin zone for the square lattice, the Mott criticality occurs independently through quite a different fluctuation, namely through the singular density fluctuations at small wavenumber.  If the antiferromagnetic order exists at low temperature of the Mott insulating phase, the antiferromagnetic fluctuations around $(\pi,\pi)$ coexist with the density fluctuations around small wavenumber arising from the Mott criticality.  
It should also be noted that this fluctuation has a completely different origin from the ordinary charge-order fluctuations at a commensurate wave vector, although the present instability may also trigger the charge ordering.

It is naturally expected that the density fluctuations at small wavenumber inherent to the Mott transition may play a novel role in stabilizing the superconducting phase.  In fact, the density fluctuations may be the origin of instabilities to various symmetry breakings including not only superconductivity but also charge and magnetic orderings, since the diverging density fluctuations are directly connected with the flattening of the quasiparticle dispersions at the Fermi level leading to the enhanced density of states.  The diverging density of states widely enhances the instability for various orders.

\begin{figure}
\begin{center}
\includegraphics[width=8cm,height=7cm]{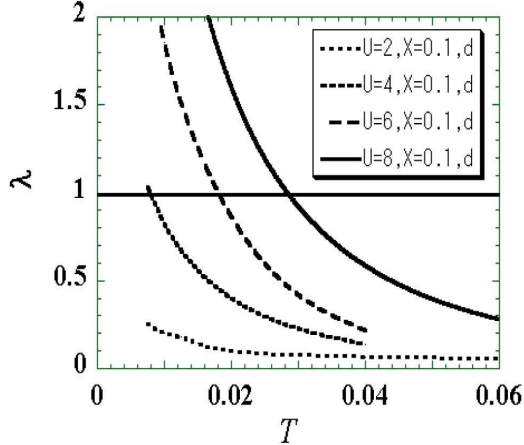}
%$$ \psboxscaled{360}{imadaFig3.eps} $$
\end{center}
\caption{
Eigenvalues $\lambda$ for the linearized Eliashberg equation (\ref{7}) are plotted as functions of temperature.  Since the eigenvalue for the $d_{x^2-y^2}$ symmetry is the largest, it is plotted for $X=0.1$. The transition temperature $T_{sc}$ is obtained from the temperature where the eigenvalue crosses $\lambda=1$ (horizontal line).   
We have used the density susceptibility in Fig.~\ref{Fig.2}. 
 We plot for several choices of the Hubbard $U$ interactions with the energy unit of the nearest-neighbor transfer $t$ on the square lattice Hubbard model.}
\label{Fig.3}
\end{figure}

The pairing mechanism arising from this density fluctuation was recently studied by a perturbative scheme of the mode-coupling theory~\cite{Imada2005}.                
We here discuss superconductivity assuming the proximity to the Mott quantum critical point in detail. In the present scope, the density fluctuations have the largest energy scale over spin and orbital fluctuations and are the primary origin of the unusual properties.  This proximity of the Mott transition indeed mediates the Cooper pairing through the enhanced density fluctuations.  When we follow the perturbative scheme, the effective interaction between two electrons is obtained from the density fluctuations as 
\begin{equation}
\Lambda(q,i\omega_n)=U-U^2\chi_X(q,i\omega_n)/2
\label{7-0}
\end{equation}
up to the second order in $U$ with $\chi_X$ obtained from (\ref{3}).  Of course other fluctuations as spin fluctuations also affect the effective interaction.  However, to extract the role of density fluctuations clearly, we ignore the contribution from spin fluctuations in this article as the first step.  In fact, the density fluctuation plays the dominant role because of its large energy scale. When spin fluctuations are also considered, we expect that it reinforces the pairing, since both of the fluctuations enhance the same type of pairing symmetry while they do not interfere each other because of their fluctuations at very different wave numbers as we see below.  We obtain the linearized Eliashberg equation for the superconducting gap $\Delta$ as
\begin{eqnarray}
\Delta(q,\omega_n)&=&-\frac{T}{N}\sum_{k,m}G(k,i\omega_m)G(-k,-i\omega_m) \nonumber \\
&&\times\Lambda(q-k,i(\omega_n-\omega_m))\Delta(k,i\omega_m),
\label{7}
\end{eqnarray}
where $N$ is the number of sites. This Eliashbrg equation is solved selfconsistently.  Considering the level of perturbative approximations here, the first nontrivial way to solve this problem is to take the bare Green's function for $G$, and we ignore the normal self-energy corrections to $G$. In the calculation of Green's function in Eq.(\ref{7}), the standard Hubbard model on the square lattice is employed as an example. However, the nonperturbative effect is taken into account through Eq.(\ref{3}) for $\chi_X$ with the parameter values introduced above. Then Eq.(\ref{7-0}) is inserted to Eq.(\ref{7}) and  the eigenvalue $\lambda$ is calculated for the right-hand side of Eq.(\ref{7}).  Namely, the linearized Eliashberg equation is solved selfconsistently for the relevant parameter values for the cuprate superconductors as cited above.  

\subsection{Unconventional pairing}
\begin{figure}
\begin{center}
\includegraphics[width=8cm,height=6cm]{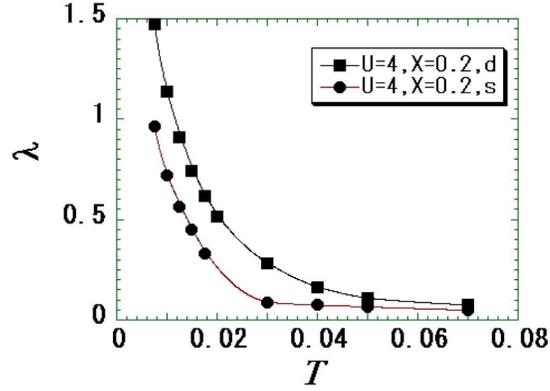}
\end{center}
\caption{Temperature dependence of the eigenvalue for the linearized Eliashberg equation for the $d_{x^2-y^2}$ (squares) and extended $s$-wave (circles) symmetries.}
\label{d-seigvl-T.eps}
\end{figure}

\begin{figure}
\begin{center}
\includegraphics[width=8cm,height=7cm]{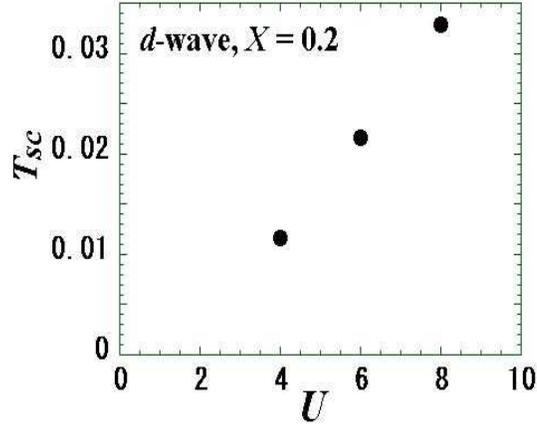}
\end{center}
\caption{$U$ dependence of the superconducting transition temperature for the $d_{x^2-y^2}$ symmetry in the energy scale of $t$. The transfer is assumed to have $t\sim 0.4$ eV and for other parameter values see the text.}
\label{Tc-Un.eps}
\end{figure}

The solution of the Eliashberg equation Eq.(\ref{7}) for the parameter values above shows that the right hand side of the linearized Eliashberg equation has the largest eigenvalue for the $d_{x^2-y^2}$  pairing symmetry, which leads to the highest superconducting transition temperature $T_{sc}$ for this pairing symmetry as we show in Fig. \ref{Fig.3}. Figure \ref{d-seigvl-T.eps} shows that the $d_{x^2-y^2}$  pairing symmetry indeed wins over the eigenvalues for the other symmetries including the symmetry of the extended $s$-wave-symmetry pairing.  It is remarkable that even though the fluctuations are at a small wavenumber $Q$, it generates the anisotropic pairing.  This is because the effective interaction is repulsive in the most part of the Brillouin zone because of the first term in Eq.~(\ref{7-0}),  while it becomes attractive only in the small wavenumber region.  This forces the pairing to have an anisotropy with nodes.  The $d_{x^2-y^2}$  symmetry and its node position are understood because the largest gap grows in the $(\pi,0)$ and $(0,\pi)$ regions, which is stabilized by the flat dispersion. Then the only possibility is to make nodes in the diagonal direction in the Brillouin zone. We note that the gap amplitude may be substantially underestimated because we have underestimated the flatness of the dispersion in Green's function by taking the bare Green's function instead of the correct one.  Figure~\ref{Fig.3} shows how the eigenvalue grows with lowering temperatures.  The superconducting transition temperature within this approximation is estimated from the temperature where the eigenvalue exceeds unity in Fig.~\ref{Fig.3}.  The transition temperature has the order of $0.01t$ to $0.05t$ as we see in Fig. \ref{Tc-Un.eps}, which corresponds to the order of 100K for the copper oxides when we take $t\sim 0.4$eV.  It should be noted that the large energy scale (namely the Mott gap scale $\sim 2$ eV) of the density fluctuation represented here by the parameter value $D_s=30X$ is crucial for achieving such a high transition temperature.  
In other words, the high-energy excitations substantially contribute to enhancing the transition temperature.  It is remarkable that within this simple approximation, the quantum Mott criticality has a dramatic effect on the superconductivity, which is comparable or even larger than that by the magnetic mechanism in the same level of approximations. 

If we properly consider the self-energy effects in Eq.(\ref{7}), we expect that the superconducting transition temperature $T_{sc}$ becomes vanishing at the Mott transition point although the pairing interaction is most enhanced at the Mott transition point because of the critical enhancement of the density fluctuations.  These two should cause the separation of $T_{sc}$ and the gap amplitude. It leads to the pseudogap behavior in the underdoped region.  This is left for future studies. We note that one has to be careful in taking account of the self-energy effects because a part of it appears through the density fluctuation itself, which is already taken into account here but is beyond the presently available perturbative treatment in the literature.

The present analyses based on the perturbative treatment and mean-field-type Eliashberg equation do not take into account fluctuation effects  particularly in two-dimensional systems.  In fact, in purely two-dimensional systems, we expect Berezinskii-Kosterlitz-Thouless (BKT) type transition for the gauge symmetry breaking of the superconductivity and  this aspect is not considered here.  Nonetheless, the present analyses have significance in the following points:  First, even when the BKT  transition is expected in pure two-dimensional system, its transition  temperature has a comparable value to the mean-field results as known in  the analyses of the XY model universality.  Therefore, the present  mechanism of the superconducting transition may also work for the BKT  transition at the similar temperature scale, which can be inferred from  the present simple approximations.  Second, the significance of the  present approach is that the superconducting mechanism arises from a  completely new origin of the proximity from the quantum Mott criticality  with the enhanced density fluctuations at small wavenumber. 
As a first step, clarification of possible relevance in the experimental situation is desired even at the mean-field level.  Third, it is useful to compare consequences of the present mechanism with the conventional ones including magnetic fluctuation mechanisms at the same level of approximations. For the mean-field analyses, it is possible because we have many available results for the conventional mechanisms in the literature.  Fourth, in the cuprates, very weak interlayer coupling, which still does not destroy the dominance of two-dimensional Mott criticality in a wide region, may sensitively induce real superconducting transition as the three-dimensional one.  This circumstance may show that  the mean-field treatment for the superconducting transition offers a  qualitatively correct way of understanding if the pairing mechanism is  correctly picked up.  More detailed analyses with consideration of the inherent two-dimensional fluctuations and the resultant BKT transition is left for future studies.

The present pairing mechanism may also work near the bandwidth-control transition point.  In fact the basic mechanism can be straightforwardly applied to the region near the critical end point of the bandwidth-control transition. This may explain the superconducting phase observed near the Mott transition point of $\kappa$-ET compounds family~\cite{Kanoda0}.  

The superconductivity near the valence instability point was studied theoretically as a model for CeCu$_2$Si$_2$, CeCu$_2$Ge$_2$, and other heavy fermion compounds~\cite{Miyake,Monthoux}. Since we expect a similar novel criticality to the present Mott criticality, it would be an intriguing issue to pursue the mechanism of the superconductivity along the same line.

Since electron density fluctuations of course strongly couple with phonon modes, the phonons also supplement this density fluctuation mechanism.  In fact, the strong coupling to small wavenumber phonon necessarily occurs, if the enhanced electron density fluctuations already exist in the Mott critical region.  This may reinforce the strong coupling to phonons even when the conventional electron-phonon coupling $\lambda$ is rather weak.  In fact, the $B_{1g}$ out-of-plane phonon mode with a small momentum transfer~\cite{Cuk} may have a relevance and presumable resultant kink structure in the angle-resolved photoemission spectra should be considered under this circumstance.

\section{Summary and Discussion}
In this paper, it has been shown that the Mott transition is successfully described by a new framework for quantum phase transitions.  The natural order parameter for the Mott transition is the electron doping concentration for the filling-control transition.  For the bandwidth-control transition, the natural order parameter is the doublon (or holon) density.  At zero temperature, the phase boundaries of metals and Mott insulators always exist as the Mott transition, which occurs either as the first-order or continuous transitions. 

If the Mott transition occurs as the continuous transition at zero temperature and other spontaneous symmetry breakings are not involved, the metals and insulators are adiabatically connected at finite temperatures. We call this regime the quantum regime surrounding the $T_c=0$ boundary.
On the other hand, if the transition occurs as the first-order transition at $T=0$ and terminates as the critical point above the Fermi degeneracy temperature, we call it the classical regime.
Sandwiched by these two regimes, the marginal quantum critical region appears, where the first-order boundary and the continuous $T_c=0$ boundary meet at $T=0$. 

Although the density fluctuation is completely suppressed in the Mott insulator because of the Mott gap, the criticality of the continuous Mott transition in the metallic side can be described by the critical enhancement of the density fluctuation at small wavenumbers in contrast to the naive expectation. This critical enhancement indeed occurs at finite-temperature Mott critical line as well as at the marginal quantum critical point. The marginal quantum criticality shows nontrivial and novel features.

In the classical regime around the high critical temperature, the Mott criticality is described by the Ising universality class. However, in the quantum regime, the clarified Mott criticality indicates the breakdown of the Ginzburg-Landau-Wilson (GLW) scheme. The quantum dynamics is derived from the path-integral formalism with the one-to-one correspondence between single-particle dynamics and two-particle correlations.  The divergence of time scale at the transition is described by the dynamical exponent $z=2$ for the quantum region and $z=4$ for the marginally quantum region.  Remarkably, it is shown that the free energy has nonanalytic expansion with respect to the order parameter, with power depending on spatial dimensionality $d$ for the quantum as well as the marginally quantum criticality, in marked contrast with the GLW expansion, which is demensionality independent.  This unusual expansion results in dimensionality-dependent critical exponents, which also indicates that the Mott transition occurs always at the upper critical dimension at any $d$.  Then, at any $d$, the scaling relations and the hyperscaling are still satisfied while the mean-field description is basically justified for the critical exponents except for the possible logarithmic corrections.  They are totally described by a new universality class.  Particularly when the Mott critical temperature becomes lowered just to zero temperature, the marginal quantum critical point appears and the critical exponents are given by $\gamma=2-d/2, \beta=d/2$ and $\delta=4/d$.  

The present theoretical framework for the quantum Mott criticality has clarified many aspects which are consistent with the experimental results near the Mott insulator.  These are summarized in the following:

First, the universality class of the quantum Mott transition well explains the otherwise puzzling critical exponents recently discovered in $\kappa$-(ET)$_2$Cu[N(CN)$_2$]Cl; $\gamma=1, \beta=1$ and $\delta=2$. This exponents are identified as those at the marginal quantum critical point in two-dimensional systems. The scaling description is completely consistent with both the classical Ising-type transition observed in V$_2$O$_3$ and the quantum transition observed in $\kappa$-(ET)$_2$Cu[N(CN)$_2$]Cl as well as in $\kappa$-(ET)$_2$Cu$_2$(CN)$_3$.  

Second, such unusual exponents and the new universality class at the marginal quantum critical point inevitably cause the differentiation of electrons in the momentum space even when the large Fermi surface with the contained Luttinger volume is expected. This draws a concrete picture how the Fermi liquid breaks down to the Mott insulator.  The differentiation along the Fermi surface is the driving mechanism of the emergence of the flat dispersion, and the arc structure observed by the angle resolved photoemission experiments in the cuprate superconductors.  The differentiation generates particular points on the Fermi surface responsible for the criticality, which is the reason why the hyperscaling relation is satisfied in the present theory.  
     
Third, approaching the marginal quantum critical point, the system becomes more and more sensitive to the external perturbations and is easily driven to the inhomogeneous state, which has been suggested in various types of surface probes.  We have estimated the typical length scale of the inhomogeneity determined from the balance of the Mott criticality and the long-range Coulomb interaction in the filling-control transition.

At the same time, the universality class of the Mott criticality at finite temperatures in the classical region is protected from the randomness because the Ising universality at finite temperatures is not influenced by the small randomness.  The first-order transition is only driven by the mechanism of the Mott transition equivalent to the Ising class. The universality class of the critical point at the termination point of the first-order transition may receive fewer effects of Anderson localization, because the Anderson transition does not drive the first-order transition at all. The insensitivity to the randomness is particularly true in three dimensions. 
In purely two dimensional systems, however, the situation is nontrivial because of the sensitive random field effects on the Ising transition, which will be discussed in a separate paper.  This is also the origin of the tendency for the spatial inhomogeneity.   At the marginal quantum critical point, we also expect that the universality is protected against randomness at least in three dimensions because of diverging density of states at the gap edge and efficient screening through divergent density susceptibility.  
Near the marginal quantum critical point, the enhanced density of states implies that the Thomas-Fermi screening length becomes short, which leads to a more efficient screening of the random potential by the fewer carriers.  Then contrary to the naive expectation, the density fluctuation at a long length scale appears.
Along the $T_c=0$ boundary, the metallic phase may be under a severe effects of randomness and the continuous metal-insulator transition is eventually triggered by the Anderson localization.  

The filling-control transition requires a special care because of the long-range Coulomb interaction.  For charged electrons, the real divergence of the density fluctuation at strictly $q=0$ is suppressed.  However, the density fluctuations are still strongly enhanced at a small nonzero wavenumber, which may cause unusual properties.  In addition, if dispersive light carriers coexist with the carriers near the Mott insulator, the screening and compensation by such dispersive and good metallic carriers allows the critical divergence of the density fluctuation for the correlated carriers. This appears to be realized indeed in the cuprate superconductors.
 
The fourth point for the experimental relevance of the marginal quantum criticality is the non-Fermi-liquid properties.  The non-Fermi-liquid properties observed in the doped Mott insulators with two-dimensional anisotropy are accounted for by the density fluctuations arising from the quantum Mott criticality.
This in fact explains the $T$-linear resistivity in two dimensions, when we employ the mode coupling theory. 
In particular, the fluctuations at the energy scale as large as 1 eV observed as the long tail in the optical conductivity universal in transition metal compounds and organic compounds are explained not by spin or orbital fluctuations but by this density fluctuation of the quantum Mott criticality.   

We have also shown that the mechanism of the high-$T_c$ superconductivity can be ascribed to the density fluctuations originating from the marginal quantum Mott criticality.  Although the density fluctuation occurs at small wave number, it causes the unconventional pairing with a nodal structure. For the realistic parameter values of the cuprate superconductors, the solution of the linearized Eliashberg equation has the highest transition temperature of the order of 100K for the $d_{x^2-y^2}$  wave symmetry .  It is remarkable that, within the present level of approximation, the density fluctuations inherent near the {\it quantum Mott criticality} overlooked in the literature cause comparable or even larger effects than the spin fluctuation mechanisms extensively studied for the cuprate superconductors. 
The large energy scale of the density (charge) fluctuations may help even stronger instability towards the superconductivity, if this instability could be more carefully tuned by the design of material parameters. This is a challenging future task. 

From the experimental point of view, it is highly desired to develop a good experimental probe for studying the dynamical and short-range density (charge) correlations.  In contrast to the magnetic correlations well studied by neutron scattering and NMR, experimental probes for the wavenumber and frequency dependent charge correlations of electrons are poor.  We can study the optical conductivity and dielectric functions only at zero wavenumber while STM and several microscopes can detect only the static structure on the surface. Raman scattering is not powerful enough so far for the study of the systematic wavenumber dependence.  Electron energy loss spectra and inelastic X-ray scattering in principle probe the dynamical density correlations while the present energy resolution is rather poor. To uncover the whole feature of physics near the Mott transition, it would be highly desired to develop experimental probes for the frequency and momentum dependence of electron density correlations and it will make a breakthrough in this field.  Our prediction from the present work is that the extended charge (density) fluctuations at small wavenumbers may be observed in the frequency and wavenumber dependent spectra in the critical region of a certain class of the Mott transitions in transition metal compounds and organic conductors. The density fluctuation may also occur with a compensation between two different types of carriers, good metallic carriers and the strongly correlated carriers near the Mott insulator. This compensated density fluctuation will require a more refined probe to be detected.   This will reveal the missing ring of various puzzling properties including the mechanism of the high-$T_c$ cuprate superconductors and the criticality of the Mott transition in the organic conductors. 

The author thanks D. Basov, K. Kanoda, F. Kagawa, S. Tajima and Y. Tokura for illuminating discussions on their experimental results.  This work is supported by the grant-in-aid from the Ministry of Education, Culture, Sports, Science and Technology. A part of the computation has been performed at Supercomputer Center, Institute for Solid State Physics, University of Tokyo.

%\bibliography{sample}

\begin{thebibliography}{10} 
\bibitem{RMP}For a review, see M. Imada, A. Fujimori and Y. Tokura, Rev. Mod. Phys. {\bf 70}, 1039 (1998).
\bibitem{PeierlsMott} N.F. Mott, in Proc. Phys. Soc. A{\bf 49}, 72 (1937). 
\bibitem{deBoer} J.H. de Boer, and E.J.W. Verway, Proc. Phys. Soc. A{\bf 49}, 59 (1937).
\bibitem{Mott} N.F. Mott, Proc. Phys. Soc. A{\bf 62}, 416 (1949).
\bibitem{Mott2} N.F. Mott, {\it Metal Insulator Transitions} (Taylor and Francis, London/Philadelphia,1990)
\bibitem{Bednorz} J.G. Bednorz and K.A. Mueller, Z. Phys. B{\bf 64}, 189 (1986). 
\bibitem{Morita} H. Morita, S. Watanabe and M. Imada, J. Phys. Soc. Jpn. {\bf 71}, 2109 (2002).
\bibitem{Kashima1} T. Kashima and M. Imada, J. Phys. Soc. Jpn. {\bf 70}, 3052 (2001).
\bibitem{KanodaSpinliquid} Y. Shimizu, K. Miyagawa, K. Kanoda, M. Maesato and G. Saito, Phys. Rev. Lett. {\bf 91}, 107001 (2003).
\bibitem{Fukuyama} K. Ishida, M. Morishita, K. Yawata and H. Fukuyama, Phys. Rev. Lett.{\bf 79}, 3451 (1997).
\bibitem{Ishimoto} R. Masutomi, Y. Karaki and H. Ishimoto, Phys. Rev. Lett. {\bf 92}, 025301 (2004). 
\bibitem{Mott1storder} N.F. Mott, Can. J. Phys. A{\bf 34}, 1356 (1956).
\bibitem{WatanabeGPIRG}S. Watanabe and M. Imada,  J. Phys. Soc. Jpn. {\bf 73}, 1251 (2004). % 
\bibitem{ImadaKashima} M. Imada and T. Kashima, J. Phys. Soc. Jpn. {\bf 69}, 2723 (2000).
\bibitem{KashimaImadaPIRG} T. Kashima and M. Imada, J. Phys. Soc. Jpn. {\bf 70}, 2287 (2001). % 
\bibitem{Fournier} D. Fournier, M. Poirier, M. Castonguay, and K. Truong, Phys. Rev. Lett. {\bf 90}, 127002 (2003).
\bibitem{Lefebvre} S. Lefebvre, P. Wzietek, S. Brown, C. Bourbonnais, D. Jerome, C. M\'ezi\`ere, M. Fourmigu\'e and P. Batail, Phys. Rev. Lett. {\bf 85}, 5420 (2000).
\bibitem{Kanoda} F. Kagawa, T. Itou, K. Miyagawa and K. Kanoda, Phys. Rev. B {\bf 69}, 064511 (2004).
\bibitem{Tokura} I. K\'ezsm\'arki, N. Hanasaki, D. Hashimoto, S. Iguchi, Y. Taguchi, S. Miyasaka and Y. Tokura, Phys. Rev. Lett. {\bf 93} 266401 (2004).
\bibitem{Kanoda3} F. Kagawa, K. Miyagawa and K. Kanoda, unpublished.
\bibitem{FurukawaImada0} N.Furukawa and M. Imada, J. Phys. Soc. Jpn. {\bf 60}, 3604 (1991).
\bibitem{FurukawaImada} N.Furukawa and M. Imada, J. Phys. Soc. Jpn. {\bf 61}, 3331 (1992).
\bibitem{FurukawaImada2}N. Furukawa and M. Imada,
  J. Phys. Soc. Jpn. {\bf 63}, 2557 (1993).
\bibitem{Imada2004} M. Imada, J. Phys. Soc. Jpn. {\bf 73}, 1851 (2004).
\bibitem{Imada2005} M. Imada, J. Phys. Soc. Jpn. {\bf 74}, 859 (2005).
%\bibitem{Imada2005} M. Imada, J. Phys. Soc. Jpn. {\bf 74}, (2005) No.3; preprint cond-mat/0411217.
\bibitem{Limelette} P. Limelette, A. Georges, D. Jerome, P. Wzietek, P. Metcalf and J.M. Honig, Science {\bf 302}, 89 (2003).
\bibitem{Castellani} C. Castellani, C. Di Castro, D. Feinberg and J. Ranninger, Phys. Rev. Lett {\bf 43}, 1957 (1979)
\bibitem{BEG} M. Blume, V.J. Emery and R.B. Griffiths, Phys. Rev. A{\bf 4}, 1071 (1971).
\bibitem{Kotliar} G. Kotliar, S. Murthy and M.J. Rozenberg, Phys. Rev. Lett. {\bf 89}, 046401 (2002).
\bibitem{Kotliar2} G. Kotliar, E. Lange and M.J. Rozenberg, Phys. Rev. Lett. {\bf 84}, 5180 (2000).
\bibitem{Goldenfeld} For example, see N. G. Goldenfeld, {\it Lectures on Phase Transitions and Renormalization Group} (Addison-Wesley, 1992).
\bibitem{Kohn} W. Kohn, Phys. Rev. A{\bf 133}, 171 (1964).
\bibitem{Fisher} M. P. A. Fisher, P.B. Weichman, G. Grinstein and D.S. Fisher, Phys Rev. B {\bf 40}, 546 (1989).
\bibitem{Imada1995} M. Imada, J. Phys. Soc. Jpn. {\bf 64}, 2954 (1995).
\bibitem{AssaadZ} M. Brunner, F.F. Assaad and A. Muramatsu, Phys. Rev. B {\bf 62}, 15480 (2000)
\bibitem{RMPX}See Chap.II.F in Ref.1. 
\bibitem{Kohno} M. Kohno, Phys. Rev. B {\bf 55}, 1435 (1997).
\bibitem{Tsunetsugu} H. Tsunetsugu and M.Imada, J. Phys. Soc. Jpn. {\bf 67}, 1864 (1998).
\bibitem{Nakano} H. Nakano and M.Imada, J. Phys. Soc. Jpn. {\bf 68}, 1458 (1999).
\bibitem{Assaad99} F. F. Assaad and M. Imada, Eur. Phys. J. B {\bf 10}, 595 (1999).
\bibitem{Assaad96} F.F. Assaad, and M. Imada, Phys. Rev. Lett.{\bf 76}, 3176 (1996).
\bibitem{25} H. Nakano and Y. Takahashi, J. Phys. Soc. Jpn. {\bf 73}, 983 (2004). 
\bibitem{7} Z.-X. Shen and D. S. Dessau, Phys. Rep. {\bf 253}, 2 (1995).
\bibitem{8} A. Ino, C. Kim, M. Nakamura, T. Yoshida, T. Mizokawa, Z.-X. Shen, A. Fujimori, T. Kakeshita, H. Eisaki and S. Uchida, Phys. Rev. B {\bf 62}, 4137 (2000).
\bibitem{Andersen} E. Pavarini, I. Dasgupta, T. Saha-Dasgupta, 
O. Jepsen, and O. K. Andersen, Phys. Rev. Lett. {\bf 87}, 47003 (2001).
\bibitem{Ino} A. Ino, C. Kim, M. Nakamura, T. Yoshida, T. Mizokawa, A. Fujimori, Z.-X. Shen, T. Kakeshita, H. Eisaki and S. Uchida, Phys. Rev. B {\bf 65}, 094504 (2002).
\bibitem{Tricritical} For a review see, I.D. Lawrie and S. Sarbach, {\it Phase Transitions and Critical Phenomena} ed. by c. Domb and J.L. Lebowitz, Vol.9, p.2.
%\bibitem{9} S.H. Pan {\it et al.}, Nature {\bf 413}, 282 (2001).
\bibitem{9} Z. Wang, J.R. Engelbrecht, S. Wang, H. Ding and S.H. Pan, Phys. Rev. B {\bf 65}, 064509 (2002);S.H. Pan, J.P. O'Neal, R.L. Badzey, C. Chamon, H. Ding, J.R. Engelbrecht, Z.Q. Wang, H. Eisaki , S. Uchida, A.K. Guptak, K.W. Ng, E.W. Hudson, K.M. Lang, J.C.Y. Davis,Nature {\bf 413}, 282 (2001).
\bibitem{10} Y. Kohsaka, T. Hanaguri, K. Kitazawa, M. Azuma, M. Takano and H. Takagi, J. Low Temp. Phys. {\bf 131}, 299 (2003); Hanaguri, C. Lupien, Y. Kohsaka, D.-H. Lee, M. Azuma, M. Takano, H. Takagi and J. C. Davis, Nature {\bf 430}, 1001 (2004).
\bibitem{11} For example, M. Uehara, S. Mori, C.H.Chen and S.-W. Cheong, Nature {\bf 399}, 560 (1999).
\bibitem{12} Y. Tokura, Colossal Magnetoresistance Oxides.(Gordon \& Breach, New York, 2000).
\bibitem{13} E. Dagotto, T. Hotta and A. Moreo, Phys. Rep. {\bf 344}, 1 (2001).
\bibitem{20} S. A. Kivelson, E. Fradkin and V. J. Emery, Nature {\bf 393}, 550 (1998).
\bibitem{21} J. Zaanen, Nature {\bf 415}, 379 (2002).
\bibitem{28} T. Moriya, Spin Fluctuations in Itinerant Electron Magnetism, (Springer, Berlin, 1985).
\bibitem{29} For example, D. van der Marel, H. J. A. Molegraaf, J. Zaanen, Z. Nussinov, F. Carbone, A. Damascelli, H. Eisaki, M. Greven, P. H. Kes and M. Li, Nature {\bf 425}, 271 (2003).
\bibitem{30} M. Dumm, S. Komiya, Y. Ando and D.N. Basov, Phys. Rev. Lett. {\bf 91}, 077004 (2003).
\bibitem{Lee} Y.S. Lee, J. Yu, J.S.Lee, T.W.Noh, T.-H.Gimm, H.-Y. Choi and C.B. Eom, Phys. Rev. B {\bf 66} 041104(R) (2002); J. S. Dodge, C. P. Weber, J. Corson, J. Orenstein, Z. Schlesinger, J. W. Reiner and M. R. Beasley, Phys. Rev. Lett. {\bf 85}, 4932 (2000).
\bibitem{Moriya} T. Moriya, Y. Takahashi and K. Ueda, J. Phys. Soc. Jpn. {\bf 59}, 2905 (1990).
\bibitem{Pines} P. Monthoux, A.V. Balatsky and D. Pines, Phys. Lett. {\bf 67}, 3448 (1991).
\bibitem{AndersonBaskaran} G. Baskaran, Z. Zou and P.W. Anderson, Solid State Commun. {\bf 63}, 973 (1987).
\bibitem{Anderson} P.W. Anderson, Science {\bf 235}, 1196 (1987).
\bibitem{Kanoda0} K. Kanoda, Physica C{\bf 282}, 299 (1997).
\bibitem{Miyake} Y. Onishi and K. Miyake, J. Phys. Soc. Jpn. {\bf 69}, 3955 (2000).
\bibitem{Monthoux} P. Monthoux and G.G. Lonzarich, Phys. Rev. B {\bf 69}, 064517 (2004).
\bibitem{Cuk} T. Cuk, F. Baumberger, D.H. Lu, N. Ingle, X.J. Zhou, H. Eisaki, N. Kaneko, Z. Hussain, T.P. Devereaux, N. Nagaosa and Z.-X. Shen, cond-mat/0403521 and references therein.
%\end{reference}
\end{thebibliography}

\end{document}